\documentclass[aps,prd,twocolumn,showpacs,preprintnumbers,nofootinbib]{revtex4}
\usepackage{graphicx}
\usepackage[FIGTOPCAP]{subfigure}
\usepackage{dcolumn}
\usepackage{bm}
\usepackage[normalem]{ulem}
 \usepackage{color}

\usepackage{amsmath}
\usepackage{epstopdf}
\usepackage{amsmath,amssymb}
\usepackage{enumerate}
\usepackage{slashed}
\usepackage{hyperref}
\usepackage{varwidth}
\usepackage{adjustbox}
\newcommand{\tr}{\ensuremath{\mathrm{Tr}}}
\usepackage{subfigure}
\usepackage{bbold}
\usepackage{array,booktabs}
\usepackage{empheq}

\newcommand{\beq}{\begin{eqnarray}}
\newcommand{\eeq}{\end{eqnarray}}

\begin{document}

\title{Higher order quark number fluctuations via imaginary chemical potentials in $N_f=2+1$ QCD}

\author{Massimo D'Elia}
\email{massimo.delia@unipi.it}
\affiliation{
Dipartimento di Fisica dell'Universit\`a
di Pisa and INFN - Sezione di Pisa,\\ Largo Pontecorvo 3, I-56127 Pisa, Italy}

\author{Giuseppe Gagliardi}\thanks{Present address:  Universit\"at Bielefeld, Fakult\"at f\"ur Physik, D-33615 Bielefeld, Germany}
\email{giuseppe@physik.uni-bielefeld.de}
\affiliation{
Dipartimento di Fisica dell'Universit\`a
di Pisa and INFN - Sezione di Pisa,\\ Largo Pontecorvo 3, I-56127 Pisa, Italy}

\author{Francesco Sanfilippo}\thanks{Present address: INFN, Sez. di Roma Tre, Via della Vasca Navale 84, I-00146 Rome, Italy}
\email{f.sanfilippo@soton.ac.uk}
\affiliation{School of Physics and Astronomy, University of Southampton, SO17 1BJ Southampton, 
United Kingdom}

\date{\today}

\begin{abstract}
We discuss analytic continuation as a tool to extract the cumulants of
the quark number fluctuations in the strongly interacting medium from
lattice QCD simulations at imaginary chemical potentials.  The method
is applied to $N_f = 2+1$ QCD, discretized with stout improved
staggered fermions, physical quark masses and the tree level Symanzik
gauge action, exploring temperatures ranging from 135 up to 350 MeV
and adopting mostly lattices with $N_t = 8$ sites in the temporal
  direction.
The method is based on a global fit of various cumulants
as a function of the imaginary chemical potentials. We show that it is
particularly convenient to consider cumulants up to order two, and
that below $T_c$ the method can be advantageous, with respect to a direct
Montecarlo sampling at $\mu = 0$, for the determination of
generalized susceptibilities of order four or higher, 
and especially for mixed susceptibilities,
for which the gain is well above one order of magnitude.  
We provide
cumulants up to order eight, which are then used to discuss
the radius of convergence of the Taylor expansion and 
the possible location of the second order critical point
at real $\mu$: no evidence for such a point is found 
in the explored range of $T$ and for chemical potentials
within present determinations of the pseudo-critical line.
\end{abstract}


\maketitle

\section{Introduction}
\label{sec:intro}

The study of the QCD phase diagram at finite temperature and density
is presently a challenging problem.  Its comprehension is particularly
relevant, from a phenomenological point of view, to the physics of
compact stars and to understand the properties of the
strongly interacting medium formed in heavy ion collisions.  One
outstanding open issue is related to the nature of the deconfinement
transition, which is known to be an analytic crossover at zero baryon
density~\cite{Aoki:2006we}; it has been suggested that it 
could become a true transition at some critical value of
the baryon chemical potential $\mu_B$, which would represent the second
order Critical Endpoint (CEP) of a line of first order transitions
existing for larger values of $\mu_{B}$.

Even if indications for the existence of a CEP are given by many
effective models,
clear evidences from the first principles, in favor or against its
existence, are still lacking.  
Indeed, at present, a direct lattice evalutation of the QCD equation of state 
at finite $\mu_B$
is hindered by the well known \textit{sign problem}: 
the Euclidean path integral
measure becomes complex in the presence of quark chemical potentials, 
making it
impossible to apply ordinary Monte Carlo algorithms based on the
interpretation of the measure as a probability density.

One way to partially overcome the problem is the Taylor expansion
technique.  Assuming analyticity around $\mu = 0$, the free energy $F$
can be expanded in powers of the quark chemical potentials. Let us
consider, for instance, the case of QCD with three flavors of quarks
$(\text{u},\text{d},\text{s})$; the expansion is given by
\beq
\label{free_energy_expansion}
F(T,V,\mu_u,\mu_d,\mu_s) &=& F(T,V,0) \ \ \ \ \ \ \ \ \ \ \ \ \ \ \ \ \ \ \ \ \ \ \ \ \ \ \ \ \ \ \ \ \nonumber
\\ &+&
VT^{4}\hspace{-0.5cm}\sum_{i+j+k=even}\hspace{-0.3cm}\frac{\chi_{ijk}(T)}{i!j!k!}\hat{\mu}_{u}^{(i)}\hat{\mu}_{d}^{(j)}\hat{\mu}_{s}^{(k)}
\eeq
where $V$ is the spatial volume, $\hat \mu_q \equiv \mu_q/T$ and odd
monomials are zero due to the symmetry under charge conjugation of the
theory at zero chemical potentials. The coefficients
\begin{equation}
\label{zero_density_susceptibilities}
\chi_{ijk}(T) =
\frac{1}{VT^{4}}\frac{\partial^{(i+j+k)}F(T,\mu)}{\partial{\hat{\mu}_{u}}^{(i)}\partial{\hat{\mu}_{d}}^{(j)}\partial{\hat{\mu}_{s}}^{(k)}}\bigg|_{\mu_u
  = \mu_d = \mu_s = 0} \,
\end{equation}
are usually known as the \emph{generalized quark number
  susceptibilities}, and can be computed by means of standard
Montecarlo algorithms at zero chemical potentials.

The interest in generalized susceptibilities $\chi_{ijk}(T,\mu)$ has
increased over the last few years: cumulants of conserved charges,
i.e. baryon number $B$, electric charge $Q$ and strangeness $S$ are
directly accessible in heavy-ion collision experiments by evaluating
their event-by-event
fluctuations~\cite{Aggarwal:2010wy,Luo:2015ewa,Adamczyk:2014fia,Adare:2015aqk}.
They can be related to the derivatives of the free energy with respect
to $\mu_B$, $\mu_Q$ and $\mu_S$, which in turn can be obtained as a
linear combination of the coefficients $\chi_{ijk}$. It has been
shown~\cite{Bazavov:2012vg,Borsanyi:2013hza,Borsanyi:2014ewa} that
comparing the experimental measure of these cumulants with lattice QCD
computations, it is possible to extract the freeze-out paramers
without relying on phenomenological models such as the Hadron
Resonances Gas Model (HRG), thus directly from the first principles of
the theory. Moreover, the cumulants represent a sensitive probe of the
possible critical behavior associated with the CEP: 
the knowledge of a large enough number of terms in the
Taylor expansion in Eq.~(\ref{free_energy_expansion}) would allow to
obtain an estimate of the radius of convergence of the series, hence
of the location of the
CEP~\cite{Allton:2002zi,Allton:2003vx,Gavai:2003mf,Gavai:2004sd,Gavai:2008zr,Datta:2012pj}.
\\

However, a direct lattice computation of the generalized
susceptibilities suffers from at least two problems, which combine to
make the numerical effort increase exponentially with the
order of the susceptibility:
\\
\\
{\em i)} The calculation of each $\chi_{ijk}(T,0)$ requires the
evalutation of a number of different terms which rapidly increases
with the order. In particular, an increasing
number of inversions of the Euclidean Dirac operator $(\slashed{D} +
m)$ is required, which represent the most time consuming part of the
computation;
\\
\\
{\em ii)} The direct sampling of non-linear susceptibilities suffers
from the so-called problem of lacking of
self-averaging~\cite{Milchev1986}: the signal-to-noise ratio of these
quantities decreases as a power law of the spatial volume $V$, with an
exponent that grows with the order of the susceptibility.  This is
essentially a consequence of the central limit theorem: the
determination of non-linear susceptibilities involves the measurement
of deviations from a Gaussian distribution, but in the large volume
limit, and away from criticality, the distribution of the quark
number, $N_q$, can be well approximated by a Gaussian of variance $V
\chi_{2}$, with deviations which are suppressed as powers of the
volume.

For instance, from Eq.~(\ref{free_energy_expansion}) it follows that
$(V\chi_{2n})$ are extensive quantities, i.e. scaling linearly with
the spatial volume, however a direct computation shows that they are
formed of a combination of terms such as, $\langle
N_q^{2n}\rangle,\langle N_q^{2(n-1)}\rangle\langle
N_q^{2}\rangle,...\langle N_q^{2}\rangle^{n}$, which in the large
volume limit grow like $(\chi_{2}V)^{n}$, up to subleading powers of
$V$.
However, it is precisely after a fine cancellation of such terms that
the subleading corrections produce the final signal, which scales like
$V$, inducing a signal-to-noise-ratio scaling as $V^{(1-n)}$. This
implies that to achieve a given statistical accuracy for the
$2n$th-order susceptibility on a certain range of volumes, the 
number of sampled gauge configurations should be increased
proportionally to $V^{2(n-1)}$.
\\

As a consequence, the computation of the Taylor series in
Eq.~(\ref{free_energy_expansion}) becomes a hard numerical challenge,
when one tries to increase the order of the expansion.  It is possible
to reduce this problem
by inserting explicit external sources directly coupled to the quark
number operators, then exploiting the fact that the dependence of
lower order cumulants on the external sources contains useful
information about the higher order cumulants: this is analogous to
determining the magnetic susceptibility of a material from an analysis
of the magnetization as a function of the external magnetic field,
rather than from measuring fluctuations at zero external field. 
In our case, the external sources to be used are the
chemical potentials. Given that real-valued chemical potentials lead
to a sign problem, one can perform numerical simulations at purely
imaginary values of them, for which the fermion determinant and the
path integral measure remain real and positive 
Then, under the same assumptions of analyticity leading to
Eq.~(\ref{free_energy_expansion}), and ensuring that the chosen set of
imaginary chemical potentials stay within the analyticity domain, one
can exploit analytic continuation to determine the non-linear
susceptibilities. In practice, one determines the generalized
susceptibilities
\begin{equation}
\label{global_fit}
\chi_{ijk}
=\sum_{\substack{l=i\\m=j\\n=k}}^{\infty}\frac{\chi_{lmn}(0,0,0) \, 
\hat{\mu}_{u}^{l-i}\hat{\mu}_{d}^{m-j}\hat{\mu}_{s}^{n-k} 
}{(l-i)!(m-j)!(n-k)!}
\end{equation}
up to a given order $i+j+k$. From a global fit of their dependence on the imaginary 
chemical potentials $\hat \mu_q = i \mu_{q,I}/T$ one can extract the 
higher order susceptibilities.

This idea has been pursued in the past, both for the case of QCD with
quark chemical
potentials~\cite{deForcrand:2002hgr,deForcrand:2003vyj,D'Elia:2002gd,D'Elia:2004at,Cea:2006yd,D'Elia:2007ke,D'Elia:2009tm,Takaishi:2010kc,Gunther:2016vcp}
and to determine the dependence of the free energy of pure gauge
theories on the topological parameter
$\theta$~\cite{Panagopoulos:2011rb,Bonati:2015sqt,Bonati:2016tvi}.
Different strategies have been chosen in the various studies. For
instance, in Refs.~\cite{D'Elia:2009tm} and \cite{Gunther:2016vcp}
only cumulants of order one have been measured (i.e. with $i+j+k = 1$ in
Eq.~(\ref{global_fit})), while up to fourth order cumulants have been
exploited in Ref.~\cite{Takaishi:2010kc} (and also in
Refs.~\cite{Bonati:2015sqt,Bonati:2016tvi} to study $\theta$
dependence); moreover, a two dimensional grid of imaginary chemical
potentials has been used in Ref.~\cite{D'Elia:2009tm} for $N_f = 2$
QCD, while a one dimensional grid, corresponding to a variation of the
baryon chemical potential $\mu_B$, has been exploited in the other
cases.  \\

The purpose of this study is to perform a systematic investigation of
this technique for the case of $N_{f} = 2+1$ QCD with physical quark
masses, aimed both at identifying 
the optimal strategy in the choice of the 
simulation points and of the measured cumulants, and at testing its
effectiveness. At the same time, we also aim at analyzing the possible
systematic effects involved in the procedure, which are mostly related
to the truncation of the series in Eq.~(\ref{global_fit}) adopted in
the fit.

To that purpose, we have performed a series of numerical simulations,
mostly for $N_{t}=8$ lattices with an aspect ratio $L/T=4$, exploring
temperatures in the range $135\, {\rm MeV} < T < 350\, {\rm MeV}$
while staying on a line of constant physics; simulations with
different aspect ratios have been performed to estimate finite volume
effects.

Numerical simulations have been performed for $O(10^2)$ different
combinations of imaginary chemical potentials for each temperature and
measuring cumulants up to order three.

In this way, we are able to determine the zero density quark number
susceptibilities, with control over truncation effects, up to the
sixth order for $T>T_{c}$ and up to the eighth order for $T<T_{c}$. A
comparison with the standard method and with results obtained in
previous studies is performed, in particular regarding the numerical
efficiency of the strategy. In the low temperature region we also try
to estimate the radius of convergence of the free energy expansion, in
order to obtain information about the possible location of the second
order critical point at real $\mu_B$.

The paper is organized as follows. In
Section~\ref{sec:Numerical_methods} we present the lattice
discretization, the strategy adopted for the choice of the simulation
points (i.e. the different combinations of imaginary chemical
potentials), and the expression of some of the computed observables.
In Section~\ref{sec:Num_result} we present our results, and finally in
Section~\ref{sec:conclusions} we draw our conclusions.

\section{Numerical methods and setup}\label{sec:Numerical_methods}

We performed simulations of $N_f=2+1$ QCD in the presence of purely
imaginary quark chemical potentials,
$\mu_f=i\mu_{f,I},\ \mu_{f,I}\in\mathbb{R}$, with $f=u,d,s$,
considering the following Euclidean partition function of the
discretized theory:
\begin{eqnarray}
  \label{partfunc}
  \mathcal{Z} &=& \int \!\mathcal{D}U \,e^{-\mathcal{S}_{(Sym.)}} \!\!\!\!\prod_{f=u,\,d,\,s} \!\!\! 
  \det{\left({M^{f}_{\textnormal{st}}[U,\mu_{f,I}]}\right)^{\frac{1}{4}}}\ \ \  \\
  \label{tlsyact}
  \mathcal{S}_{(Sym.)}&=& - \frac{\beta}{3}\sum_{i, \mu \neq \nu} \left( \frac{5}{6}
  W^{1\!\times \! 1}_{i;\,\mu\nu} -
  \frac{1}{12} W^{1\!\times \! 2}_{i;\,\mu\nu} \right), \\
  \label{fermmatrix}
  (M^f_{\textnormal{st}})_{i,\,j}&=&am_f
  \delta_{i,\,j}+\!\!\sum_{\nu=1}^{4}\frac{\eta_{i;\,\nu}}{2}\nonumber
  \left[e^{i a
      \mu_{f,I}\delta_{\nu,4}}U^{(2)}_{i;\,\nu}\delta_{i,j-\hat{\nu}}
    \right. \nonumber\\ &-&\left. e^{-i a
      \mu_{f,I}\delta_{\nu,4}}U^{(2)\dagger}_{i-\hat\nu;\,\nu}\delta_{i,j+\hat\nu}
    \right] \, .
\end{eqnarray}
$S_{(Sym.)}$ is the tree-level Symanzik action introduced in
Refs.~\cite{Weisz:1982zw,Curci:1983an}, with $W^{n\!\times \!
  m}_{i;\,\mu\nu}$ being the trace of the $n\times m$ loop in the
$(\mu,\nu)$ plane starting from site $i$.  In order to reduce UV
cutoff effects and taste symmetry violations, the staggered fermion
matrix $M^{f}_{st}$ is built up in terms of the twice stout-smeared
links $U^{(2)}_{i;\,\nu}$, which are constructed following the
procedure described in Ref.~\cite{Morningstar:2003gk} and using an
isotropic smearing parameters $\rho=0.15$.  As usual for finite $T$
simulations, periodic (antiperiodic) boundary conditions (b.c.)  are
taken for bosonic (fermionic) fields in the temporal direction, and
periodic b.c. for all fields in the spatial directions.

For each flavor, we introduce the chemical potentials following the
prescription of Ref.~\cite{Hasenfratz:1983ba}, i.e. by multiplying, in
the fermion matrix, all the temporal links in the forward (backward)
direction by $e^{+ i a \mu_{f,I}}$ ($e^{- i a \mu_{f,I}}$) (see
Refs.~\cite{Gavai:1985ie,Gavai:2014lia} for alternative discretizations). The chemical potentials coupled to quark number
operators can be conveniently rewritten in terms of those coupled to
the conserved charges, $B$, $Q$ and $S$, the relation being
\begin{eqnarray}
\label{defchem}
\mu_u &=& \mu_B/3 + 2 \mu_Q/3 \nonumber \\
\mu_d &=& \mu_B/3 - \mu_Q/3 \\
\mu_s &=& \mu_B/3 - \mu_Q/3 -\mu_S \, .\nonumber
\end{eqnarray}
As usual for staggered fermions simulations, the residual fourth
degeneracy of the lattice Dirac operator is removed by the rooting
procedure.  The Rational Hybrid Monte-Carlo
algorithm~\cite{Clark:2004cp, Clark:2006fx, Clark:2006wp} has been
used to sample gauge configurations according to Eq.~(\ref{partfunc}).

\subsection{Physical observables}

The observables measured during each simulation run correspond to the
generalized susceptibilities $\chi_{ijk}(T,\mu)$ appearing in
Eq.~(\ref{global_fit}).  In particular, we have considered all
possible combinations with $i+j+k \leq 2$ for each simulation, and
also the combinations with $i+j+k=3$ in some cases. Their explicit
lattice version reads (we limit ourselves to the second order):
\begin{eqnarray}
\label{density_susceptibility}
\chi_{1}^{f} & \equiv & \frac{N_{t}}{4N_{s}^{3}}\left\langle \tr\left(
\left(M^{f}_{st}\right)^{-1}\frac{\partial M^{f}_{st}}{\partial
  \mu_{f}}\right)\right\rangle \nonumber \\\\ \chi_{2}^{f}& \equiv &
\frac{N_{t}}{N_{s}^{3}}\left(\frac{1}{4}\right)^2
\left\langle\left[\tr\left(\left(M_{st}^{f}\right)^{-1}\frac{\partial
    M_{st}^{f}}{\partial\hat{\mu}_{f}}\right)\right]^2\right\rangle
\nonumber \\ &-&\frac{N_{t}}{N_{s}^{3}}\left(\frac{1}{4}\right)^2
\left\langle \tr\left( \left(M_{st}^{f}\right)^{-1}\frac{\partial
  M_{st}^{f}}{\partial \hat{\mu}_{f}}\right)\right\rangle^2 \nonumber
\\ &+&
\frac{N_{t}}{4N_{s}^{3}}\left\langle\tr\left(\left(M_{st}^{f}\right)^{-1}\frac{\partial^2
  M_{st}^{f}}{\partial \hat{\mu}_{f}^2}\right)\right\rangle \nonumber
\\ &-&\frac{N_{t}}{4N_{s}^{3}}\left\langle
\tr\left(\left(M_{st}^{f}\right)^{-1}\frac{\partial
  M_{st}^{f}}{\partial\hat{\mu}_{f}}\left(M_{st}^{f}\right)^{-1}\frac{\partial
  M_{st}^{f}}{\partial\hat{\mu}_{f}}\right) \right\rangle \nonumber
\\\\ \chi_{2}^{i,j} &\equiv &
\frac{N_{t}}{N_{s}^{3}}\left(\frac{1}{4}\right)^2
\left\langle\prod_{f=i,j}\left[\tr\left(\left(M_{st}^{f}\right)^{-1}\frac{\partial
    M_{st}^{f}}{\partial\hat{\mu}_{f}}\right)\right]\right\rangle
\nonumber
\\ &-&\frac{N_{t}}{N_{s}^{3}}\left(\frac{1}{4}\right)^2\prod_{f=i,j}\langle
\tr\left( \left(M_{st}^{f}\right)^{-1}\frac{\partial
  M_{st}^{f}}{\partial \hat{\mu}_{f}}\right)\rangle
\end{eqnarray}
where the presence of the factor $1/4$ is due to our staggered
discretization. Their determination requires the evaluation of the
following traces:
\begin{eqnarray} 
&&\tr\left[ \left(M_{st}^{f}\right)^{-1}\frac{\partial
      M_{st}^{f}}{\partial \mu_{f}}\right] \nonumber \\ &&\tr\left[
    \left(M_{st}^{f}\right)^{-1}\frac{\partial M^{f}_{st}}{\partial
      \mu_{f}}\right]^{2} \nonumber \\ &&\tr\left[
    \left(M_{st}^{f}\right)^{-1}\frac{\partial^{(2)}
      M^{f}_{st}}{\partial \mu_{f}^{2}}\right] \, .
\end{eqnarray}
This is has been done, as usual, by means of noisy estimators,
using 256 Gaussian random
sources for each flavour.
Confidence
intervals and bias-subtraction for non-linear estimators of
susceptibilities have been performed by means of a Jackknife
analysis~\cite{efron}.

\subsection{Choice of the simulation runs}

At fixed $N_t$, the temperature $T = 1/(N_t a)$ has been varied by
tuning the bare coupling $\beta$ and the bare quark masses $m_s$ and
$m_u = m_d = m_{l}$, so as to change the lattice spacing $a$ while
staying on a line of constant physics, with $m_{\pi}\simeq
135\,\mathrm{MeV}$ and $m_s/m_{l}=28.15$. This line has been
determined by a spline interpolation of the determinations reported in
Refs.~\cite{Aoki:2009sc,Borsanyi:2010cj,Borsanyi:2013bia}.

For each temperature, the different combinations of imaginary quark
chemical potentials have been chosen according to the following
considerations. Our strategy is to obtain information about
generalized susceptibilities at zero chemical potentials from the
dependence on $\mu_u, \mu_d$ and $\mu_s$ of the measured lower order
susceptibilities described in the previous subsection. To that aim,
in general we employ a truncated polynomial version of
Eq.~(\ref{global_fit}), i.e.
\begin{equation}
\label{global_fit2}
\chi_{ijk}  
=\sum_{\substack{l=i\\m=j\\n=k}}^{l+m+n\leq p}\frac{\chi_{lmn}(0,0,0)\,
\hat{\mu}_{u}^{l-i}\hat{\mu}_{d}^{m-j}\hat{\mu}_{s}^{n-k} }{(l-i)!(m-j)!(n-k)!}
\end{equation}
where the parameter $p$ fixes the maximum order we would 
like to determine.

The set of simulations points must contain values of $\mu_u$, $\mu_d$
and $\mu_s$ large enough, in order to be sensible to higher order
contributions. However, small values of the chemical potentials are
important as well, in order to be able to check systematics related to
truncation effects.  Therefore a reasonable choice seems to take their
values equally spaced between zero and some maximum reference value
$\mu_{max}$. This choice will be further discussed in 
Section~\ref{sub-efficiency}.

In Ref.~\cite{D'Elia:2009tm}, a two-dimensional grid of equally spaced 
chemical potentials was considered for the case of $N_f = 2$ QCD.
In this case, considering a three-dimensional grid of equally spaced chemical
potentials is surely not feasibile, since that would 
imply a number of different simulation runs scaling as 
$\mu_{max}^3$ and reaching easily
$O(10^3)$ for each temperature. Instead, we decided to fix
the simulations points along well defined trajectories 
in the three-dimensional parameter space, in particular 
we did the following choices
\beq
(\mu_u,\ \mu_d,\ \mu_s) &=& (i \mu_I,\ \ \ \ 0,\ \ \ \ 0) \nonumber \\
(\mu_u,\ \mu_d,\ \mu_s) &=& (\ \ \ 0,\ \ \ \  0,\ i \mu_I) \nonumber \\
(\mu_u,\ \mu_d,\ \mu_s) &=& (i \mu_I,\ \ i \mu_I,\ \ \ 0) \nonumber \\
(\mu_u,\ \mu_d,\ \mu_s) &=& (i \mu_I,-i \mu_I,\ \ \ 0) \label{chosenlines}
\\
(\mu_u,\ \mu_d,\ \mu_s) &=& (i \mu_I,\ \ i \mu_I ,\ i \mu_I) \nonumber \\
(\mu_u,\ \mu_d,\ \mu_s) &=& (i \mu_I,-i \mu_I,\ i \mu_I) \nonumber 
\eeq
where $\mu_I$ parametrizes each of the six different lines, with
simulation points taken with a step size $\Delta \mu_I = 0.025\ \pi T$
between zero and a maximum value $\mu_{I,max}$, which is the same for
all the lines at a given temperature $T$. In this way, keeping the
number of lines fixed, the computational effort scales linearly with
$\mu_{I,max}$.

Another aspect to be considered is whether the number of simulation
points and measured observables is large enough to fix all generalized
susceptibilities at a given order. Indeed, the number of independent
generalized susceptibilities grows rapidly with the order: for $N_f =
2+1$ it is easy to prove, exploiting the symmetry $\chi_{lmn} =
\chi_{mln}$ due to the up and down quark mass degeneracy, that at
order $N$ such number is given by $(N/2 + 1)^2$. In general, the set
of equations (\ref{global_fit2}), which are used in the global fit,
will involve only some linear combinations of such generalized
susceptibilities, which depend on the number of lines in
Eq.~(\ref{chosenlines}) and on the number of measured observables.
Those linear combinations, for each order $N$ of generalized
susceptibilities, define a matrix $A_N$, whose rank must be at least
equal to $(N/2 + 1)^2$ in order for the global fit to provide information
about all independent susceptibilities.

\begin{figure}[t!]
\centering
\includegraphics[width=0.95\columnwidth]{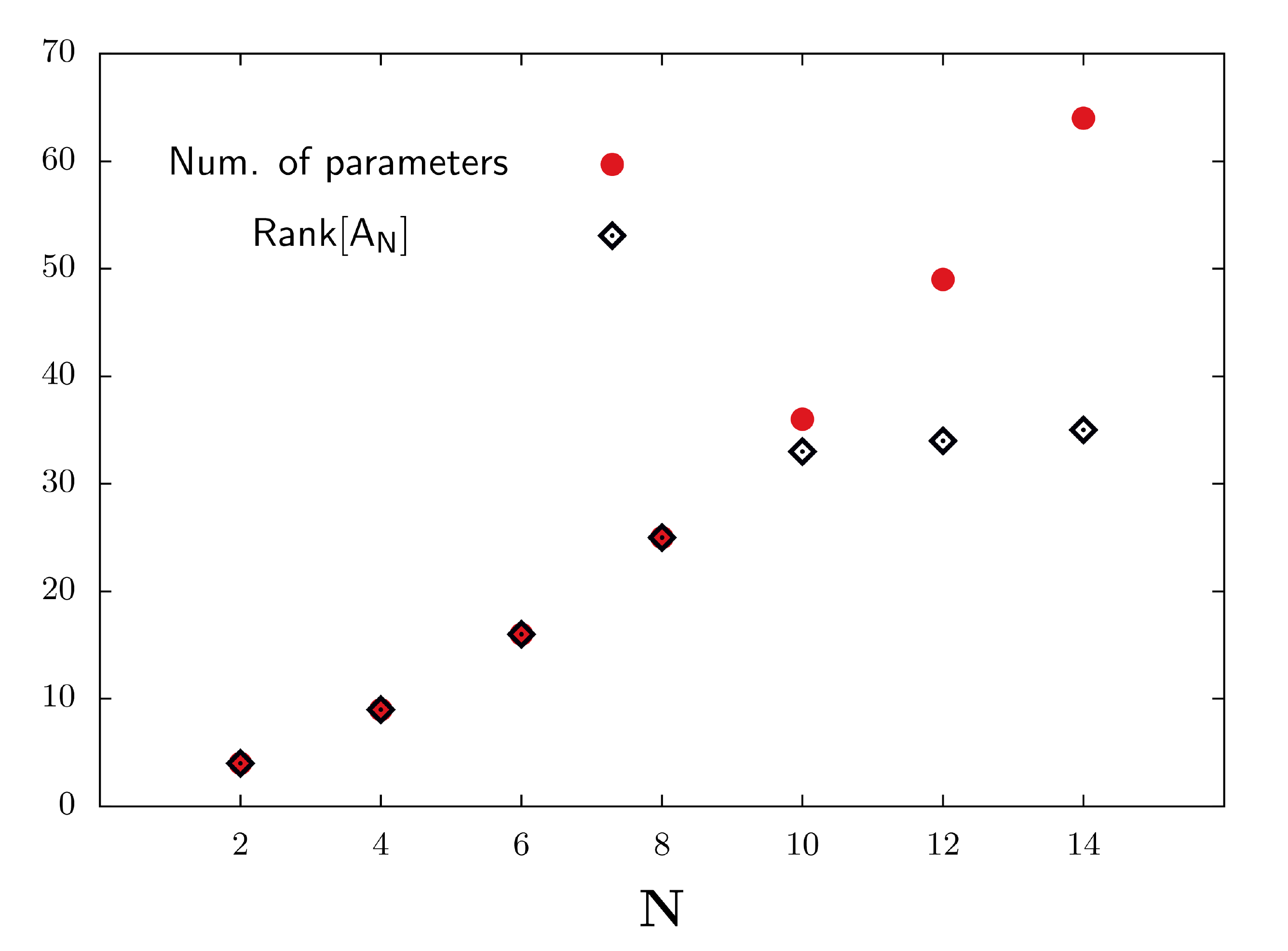}
\caption{Ranks of the $A_{N}$ matrices for some values of the order
  $N$, for simulation points chosen along the lines described in
  Eq.~(\ref{chosenlines}) and measuring as an input both quark number
  densities and second order susceptibilities. Red circles correspond
  to the number of independent $\chi_{ijk}(T,\mu=0)$ at a given order.
}\label{rank}
\end{figure}

In Fig.~\ref{rank} we show the rank of $A_{N}$ in our setup
(i.e. performing simulations along the six lines described in
Eq.~(\ref{chosenlines})), and assuming one measures all
susceptibilities up to the second order, which has been our minimal
choice for all temperatures.  It is clear that our choice suffices to
determine all the susceptibilities up to order eight, while only
$33/36$ generalized susceptibilities can be determined at order ten.

\begin{figure}[h!]
\centering
\includegraphics[width=0.95\columnwidth]{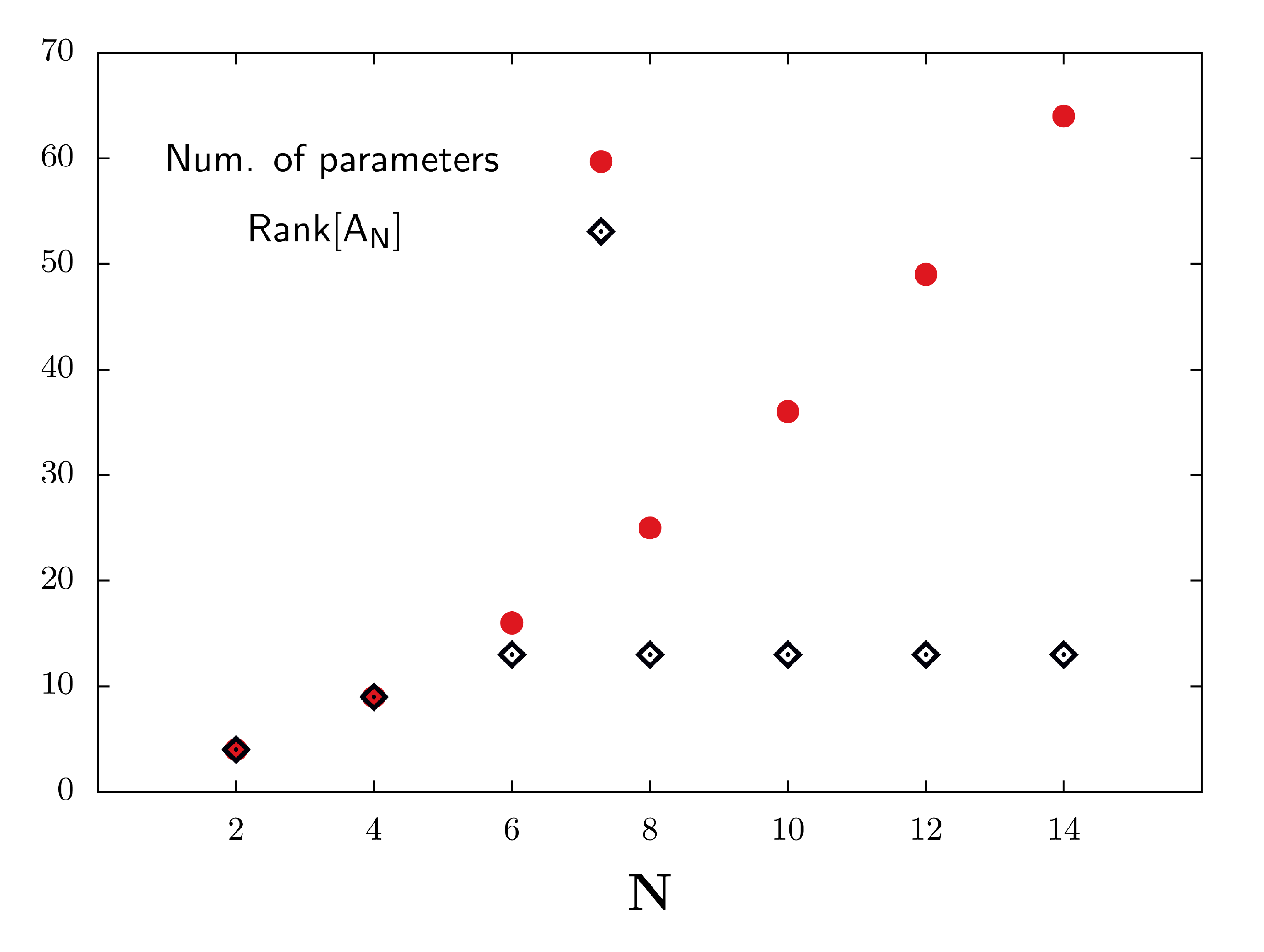}
\caption{Same as in Fig.~\ref{rank}, but with just quark number densities
  taken as an input.}\label{rank1}
\end{figure}

\begin{figure}[t!]
\centering
\includegraphics[width=0.95\columnwidth]{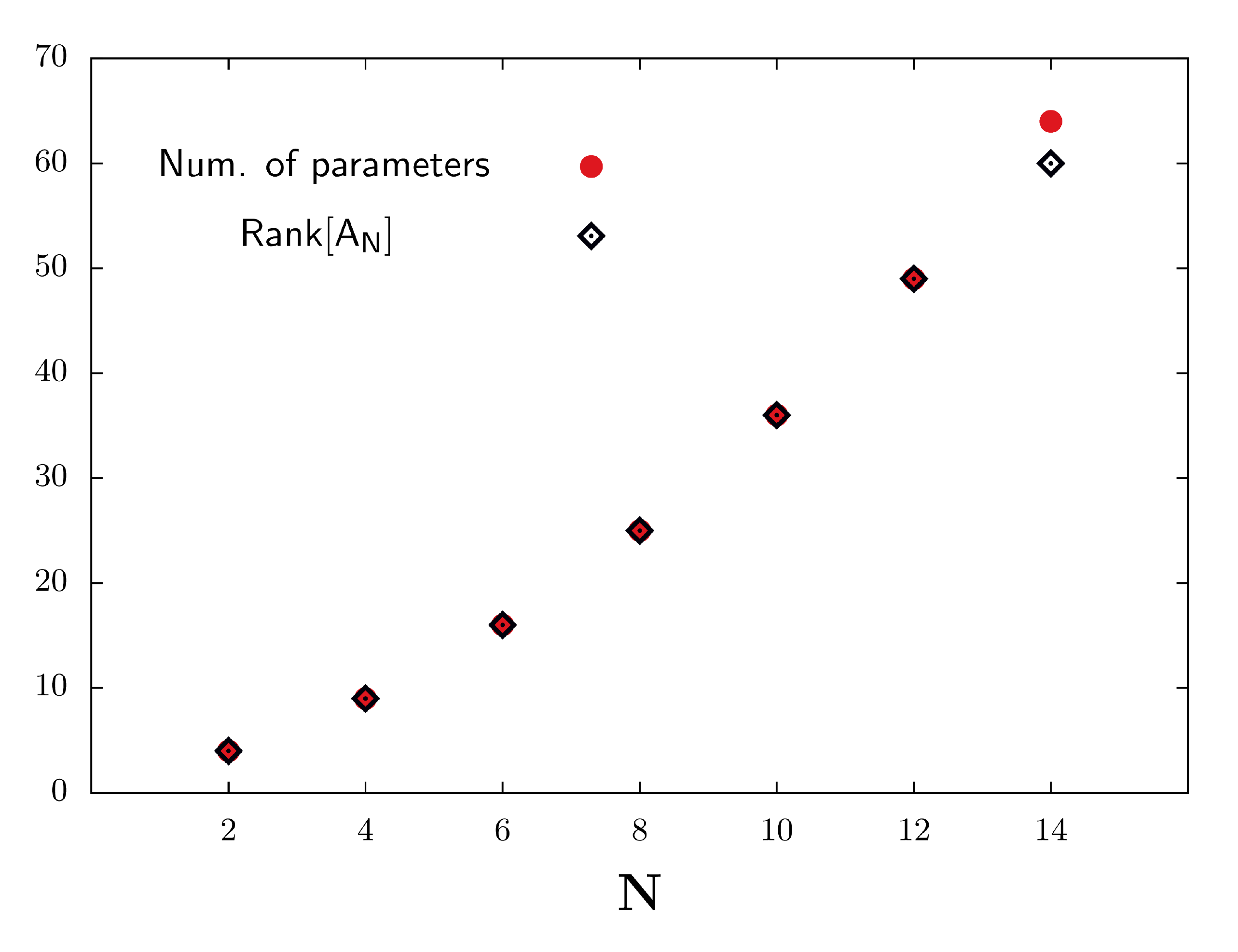}
\caption{Same as in Fig.~\ref{rank}, apart from the fact that also
  quark number susceptibilities of order three are measured and added
  as an input.}\label{rank2}
\end{figure}

\begin{figure}[h!]
\centering
\includegraphics[width=0.95\columnwidth]{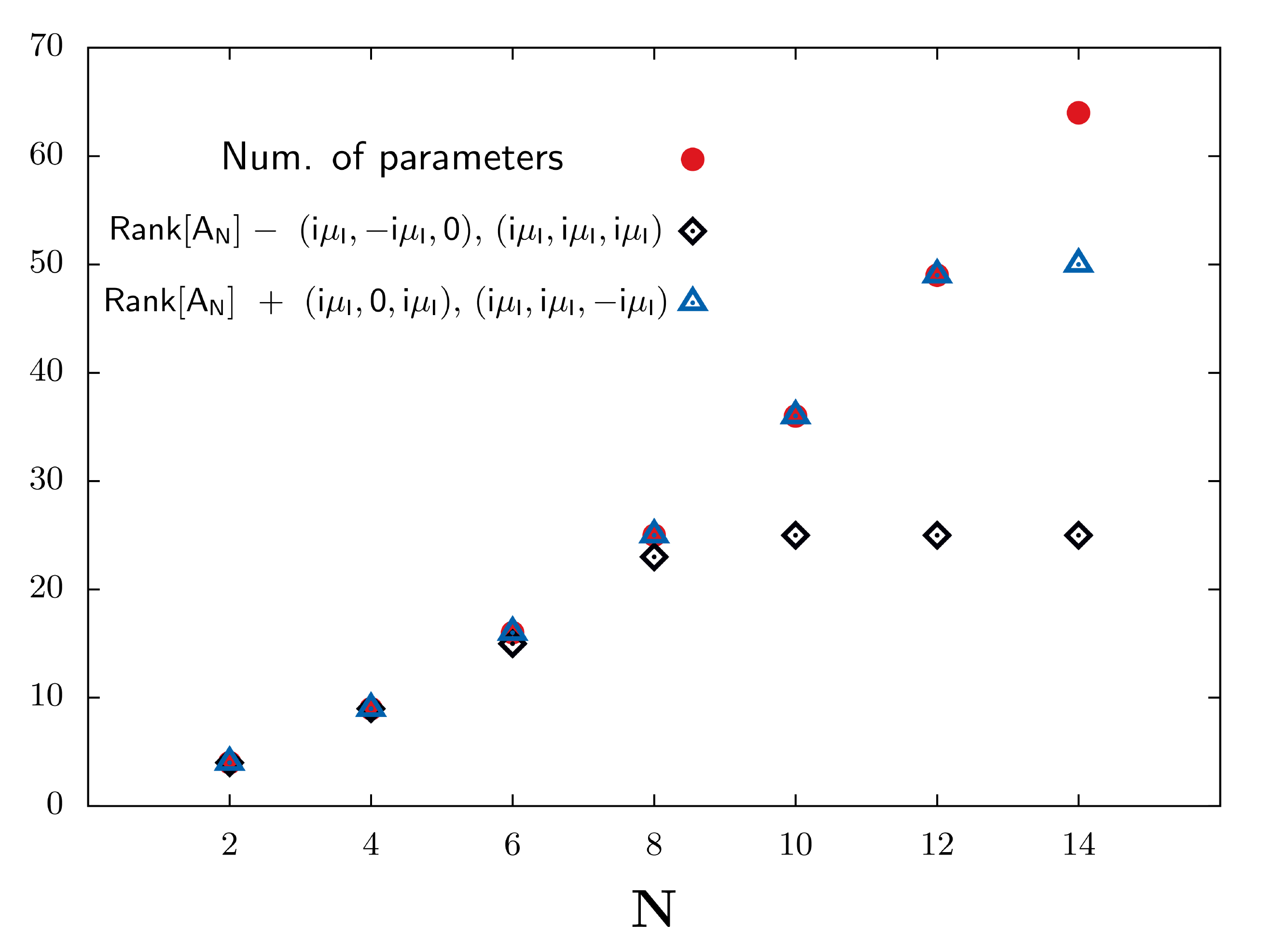}
\caption{Same as in Fig.~\ref{rank}, but considering less or more
  lines of simulation points. Triangles correspond to the case in
  which two additional lines, $(i \mu_I, 0, i \mu_I)$ and $(i \mu_I, i
  \mu_I ,- i \mu_I)$, have been added, while diamonds to the case in
  which two lines, $(i \mu_I,-i \mu_I ,0)$ and $(i \mu_I ,i \mu_I, i
  \mu_I)$, are not considered.}\label{rank3}
\end{figure}

The situation can be improved (or worsened) by changing either the
order of the measured susceptibilities, or the number of lines where
simulations are performed.  For instance, in Figs.~\ref{rank1} and
\ref{rank2} we report the same information as in Fig.~\ref{rank},
respectively for the case in which only quark number densities are
measured, and for the case in which the observables are extended up to
susceptibilities of order three (keeping the number of lines fixed).
Instead, in Fig.~\ref{rank3}, we consider the case in which one still
measures up to second order susceptibilities, but considers less or
more lines of simulation points.

The na\"ive message would seem to measure more and more
susceptibilities keeping the number of lines fixed, in order to avoid
too many simulations. However, as we have already discussed, the
precision on the observables degrades rapidly with the order, 
so what is the optimal strategy
is non-trivial and will be discussed in the following, based on
numerical results.

\begin{figure}
\centering
\includegraphics[width=0.95\columnwidth]{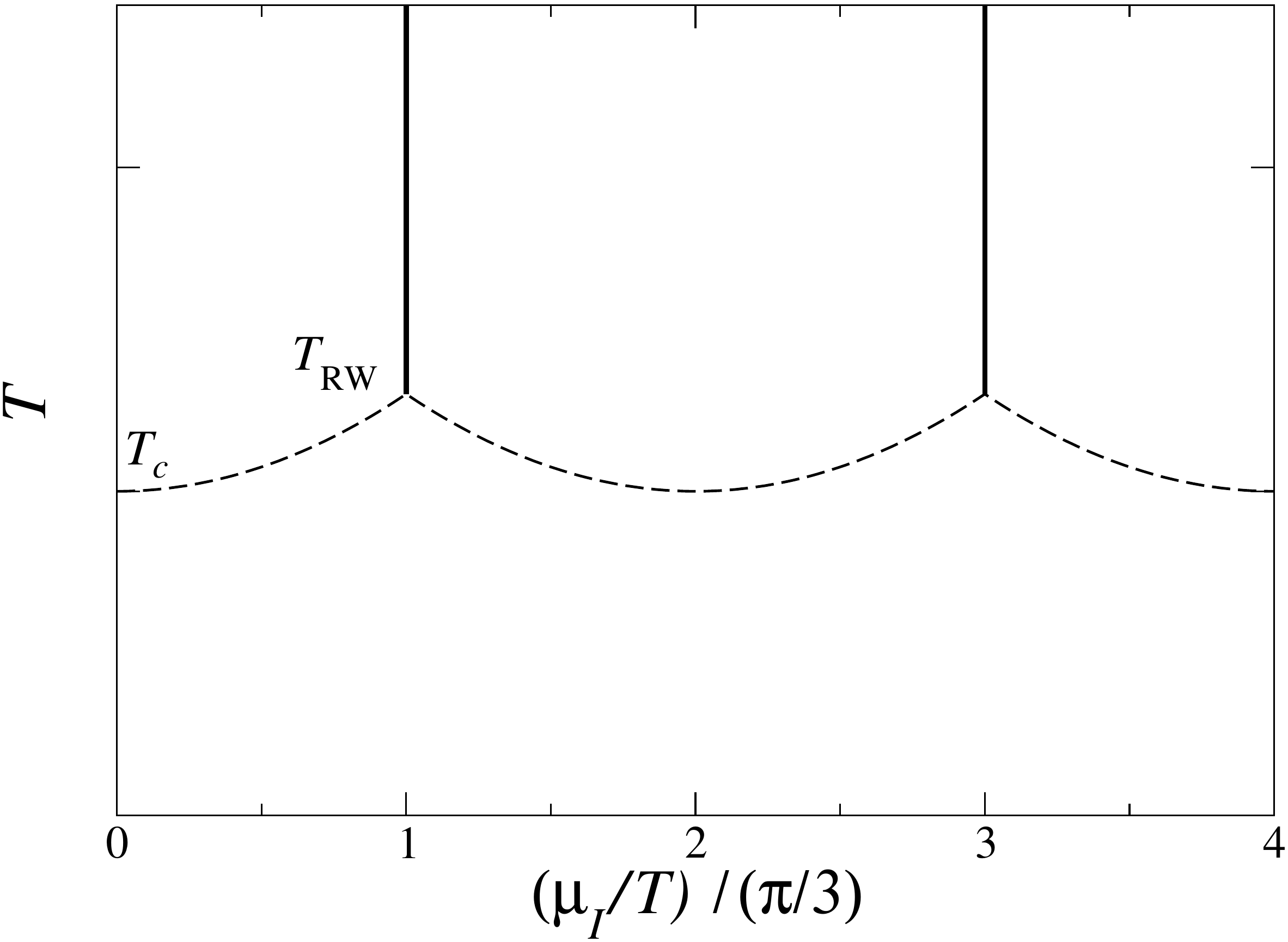}
\caption{Sketch of the phase diagram in the $T-\mu_{I}$ plane. In this
  case ($\mu_{u,d,s} = i \mu_I$) RW lines are exactly vertical and
  located at $\mu_{I}/T=(2n+1)\pi/3$. Solid lines represent first
  order phase transitions separating sectors with different
  orientation of the Polyakov loop while dashed lines correspond to
  the analytic continuation of the pseudocritical line. }\label{fig_2}
\end{figure}

There is no special limitation on the choice of $\mu_{I,max}$ in the
low temperature regime, $T < T_c$, where the partition function is an
analytic function in all chemical potentials.  In the high-$T$ region,
instead, the range of available chemical potentials is limited by the
presence of Roberge-Weiss (RW) or RW-like phase transitions,
associated to a sudden change of the expectation value of the Polyakov
loop, related to different realization of the center symmetry
breaking.  The genuine RW transition~\cite{ROBERGE1986734} is met
when moving along the line $\mu_u = \mu_d = \mu_s = i \mu_I$, which
corresponds to a pure baryon chemical potential $\mu_B = 3 i \mu_I$
(i.e. $\mu_Q = \mu_S = 0$, see Eq.(\ref{defchem})): a purely imaginary
$\mu_B$ corresponds to a global rotation of the temporal boundary
conditions for fermion fields, leading to a rotation of the fermion
contribution to the effective potential of the Polyakov loop and to a
sudden change of the global minimum for
\begin{equation} 
\mu_I/T= (2n+1)\pi /3 
\label{rwcondition}
\end{equation}
where $n$ is a relative integer. The phase diagram in the $T - \mu_I$
looks as in Fig.~\ref{fig_2}: the vertical lines are first order RW
transition lines, they start from a critical temperature $T_{RW} >
T_c$, whose value is about 200 MeV for $N_t = 8$ and about 208 MeV
in the continuum limit~\cite{Bonati:2016pwz}. The first RW line limits
the region which is analytically connected to the points at zero
chemical potentials, hence one has to take $\mu_{I,max}/T < \pi/3$ for
$T > T_{RW}$.

For intermediate temperatures, $T_c < T < T_{RW}$, no critical points
are expected, since the analytic continuation of the chiral transition
line (dashed curve in Fig.~\ref{fig_2}) is only
pseudo-critical. However one expects that, when crossing this
pseudocritical line, the dependence of the free energy 
$F$ on the chemical potentials
may become less smooth, so that systematic effects due to truncation
may become more severe, resulting in an effective limitation of the
explorable range of $\mu$.

Similar dynamics take place along the other lines, i.e.  in the more
general case $\mu_{Q},\mu_{S}\ne 0$: in those cases the exact position
of the first RW-like line depends on the quark masses and on $T$,
however one can safely state that it will occur for 
$\mu_I/T > \pi/3$, since in this case 
the different flavors tend
to orient the Polyakov loop along different directions
in the complex plane (see Refs.~\cite{Cea:2009ba,Bonati:2014rfa} for a more 
detailed discussion about this point).
For instance, along the line $\mu_u = \mu_d = i \mu_I$ and $\mu_s = 0$, it 
occurs for $\mu_I/T \sim 0.45~\pi$~\cite{Bonati:2014rfa}.

Following the discussion above, we have taken
$\mu_{I,max}/T < \pi/3$ for all temperatures above $T_c$.
Around $T_c$, the range of chemical potentials actually used in the 
global fit will be decided on the basis of the quality 
and of the stability of the fit itself.

\section{Numerical Results} \label{sec:Num_result}

Most simulations have been performed on a $32^{3}\times 8 $ lattice
for various temperatures: the complete list, including the values of
the bare parameters, is reported in Table~\ref{tab_1}. For each run,
1500 trajectories of unitary length have been performed. We measured
susceptibilities on configurations separated by 10 trajectories to
reduce autocorrelation effects\footnote{To check that our choice was 
reasonable, we measured the autocorrelation times of some typical observables. 
Throughout the explored range of temperatures, the autocorrelation times of the plaquette 
and of the quark number densities is 4-6 trajectories, 
whereas it is of O(10) trajectories 
for the chiral condensate.}. A few additional simulations have been
performed on lattices with different aspect ratios, both below and
above $T_c$, to check for finite size effects.

\begin{table}
\renewcommand{\arraystretch}{1.3}
\centering
\begin{tabular}{|c|c|c|c||c|}
\hline
\hline
$T$ [MeV] & $\beta$ & $m_{l}$ & $m_{s}$ & $\mu_{I,max}/T$ \\
\hline
\hline
135 & 3.61 & 0.002831 & 0.07971 &   $0.8\pi$  \\
\hline 
143 & 3.63 & 0.002621 & 0.07378 &  $0.8\pi$ \\
\hline
149 & 3.645 &  0.002479 & 0.06978  &  $0.4\pi$ \\
\hline
155 & 3.66 & 0.002350 &  0.06614  & $0.4\pi$ \\
\hline
160 & 3.67 & 0.002270 & 0.0639 &  $0.3\pi$  \\
\hline 
170 & 3.69 & 0.002126 & 0.5984 &  $0.3\pi$ \\
\hline
200 & 3.755 & 0.001763 & 0.04963 & $0.3\pi$\\
\hline
230 & 3.815 & 0.001516 & 0.04267 & $0.3\pi$ \\
\hline
260 & 3.87 & 0.001341 & 0.03775 &  $0.3\pi$ \\
\hline
300 & 3.94 & 0.001168 & 0.3287 &  $0.3\pi$ \\
\hline
350 & 4.0225 & 0.0009920 & 0.2792  & $0.3\pi$ \\
\hline
\end{tabular}
\caption{List of simulated temperatures and associated values of
  $\beta,m_{l},m_{s}$; $\mu_{I,max}$ represents the maximum value of
  the imaginary chemical potential used in the simulations. The value
  of $T$ is affected by an uncertainty related to the determination of
  the physical scale, which for the discretization adopted in our
  study is of the order of
  2-3~\%~\cite{Aoki:2009sc,Borsanyi:2010cj,Borsanyi:2013bia}.}
\label{tab_1}  
\end{table}

\begin{figure*}
\centering
\scalebox{0.63}{\includegraphics{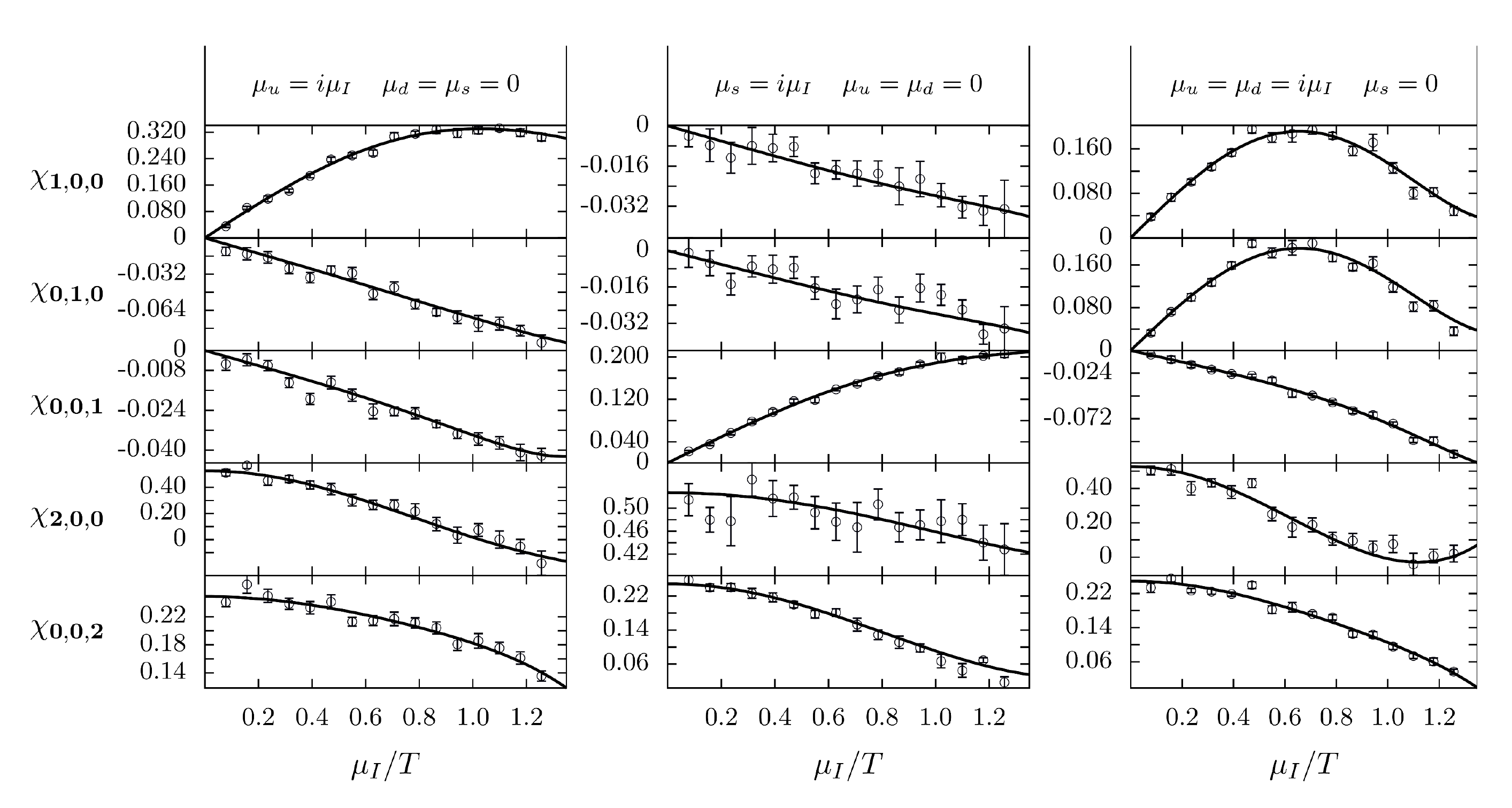}}
\caption{Example of the global fit for $T = 149$ MeV. We show only a
  subsample of a total of 54 polynomial fits which are performed at
  the same time (3 densities plus 6 second order susceptibilites
  fitted along 6 different trajectories). The best fit functions are
  taken according to Eq.~(\ref{global_fit2}) with a truncation to
  order eight. The reduced $\tilde \chi^2$ is 1.3. Notice that in the global fit we did not take into account cross-correlations between susceptibilities measured at the same chemical potential, hence the  covariance matrix has a simple diagonal form.}
\label{globalfitexample}
\end{figure*}

As outlined above, our strategy, for each temperature, has been to
perform a global fit, according to Eq.~(\ref{global_fit2}), of the
dependence on the chemical potentials of all quark number densities
and susceptibilities up to order two, along the trajectories in the
three-dimensional chemical potential space described in
Eq.~(\ref{chosenlines}). A subsample of such global fit is reported in
Fig.~\ref{globalfitexample} for $T = 149$ MeV, where we show some of
the best fit polynomials obtained according to Eq.~(\ref{global_fit2})
with a truncation $p = 8$.

\subsection{Analysis of systematic errors}

The main source of systematic error, in the analytic continuation
method, comes from the ambiguity in the choice of the fitting
function. In our case this means that coefficients resulting from the
global fit procedure, i.e. the generalized susceptibilities
$\chi_{ijk}(T)$, may depend on the order of the polynomial (i.e. on
the truncation order) as well as on the fitting range.
 
As a general procedure to keep this systematic error under control, we
started with ranges of $\mu_I$, going from zero up to a maximum value
$\bar \mu_I$, small enough so that a lowest order polynomial could
provide a good description of the data.
Next, we increased the upper value of the range, $\bar\mu_I$, keeping
the polynomial degree fixed, as long as reasonable values of reduced
chi-squared test, $\tilde \chi^{2}$, were obtained.
Otherwise, the polynomial order was increased in order to go back to
$\tilde \chi^2 \simeq 1$: at this stage, the stability of the
previously determined coefficients was checked, and any variation
going beyond the statistical errors (obtained in the global best-fit procedure) was added as a systematic error to
the final determination.

\begin{figure}
\centering
\includegraphics[width=0.95\columnwidth]{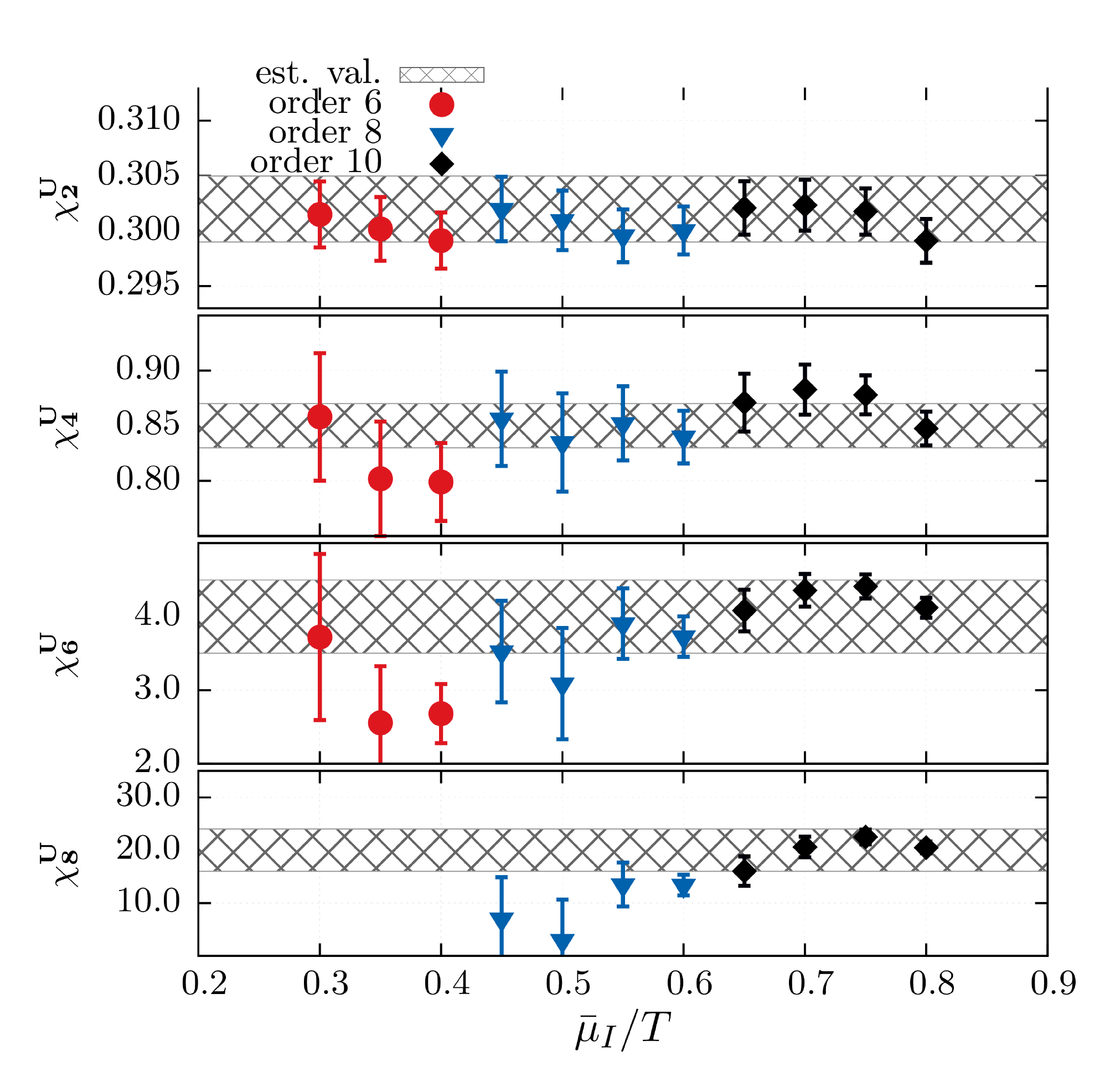} \\
\caption{An example of the procedure followed to determine systematic
  errors. The errorbars represent the statistical error obtained in the global best fit. The polynomial degree is increased every time that the
  global best fit yields non acceptable values of the reduced
  chisquared $\tilde \chi^2$. Circles, triangles and diamonds refer to
  a global fit performed with a polynomial of order 6, 8, 10,
  respectively, while the grey bands represent the final
  estimate. Data refer to simulations on the $32^{3} \times 8$ lattice
  at $T = 135$ MeV. }
\label{sistematico_err}
\end{figure}

An example of this procedure is reported in
Fig.~\ref{sistematico_err}, where we show the evolution of some
susceptibilities as the fit range or the polynomial order is changed,
for $T = 135$.

\begin{figure}
\centering
\begin{flushright}
\includegraphics[width=0.95\columnwidth]{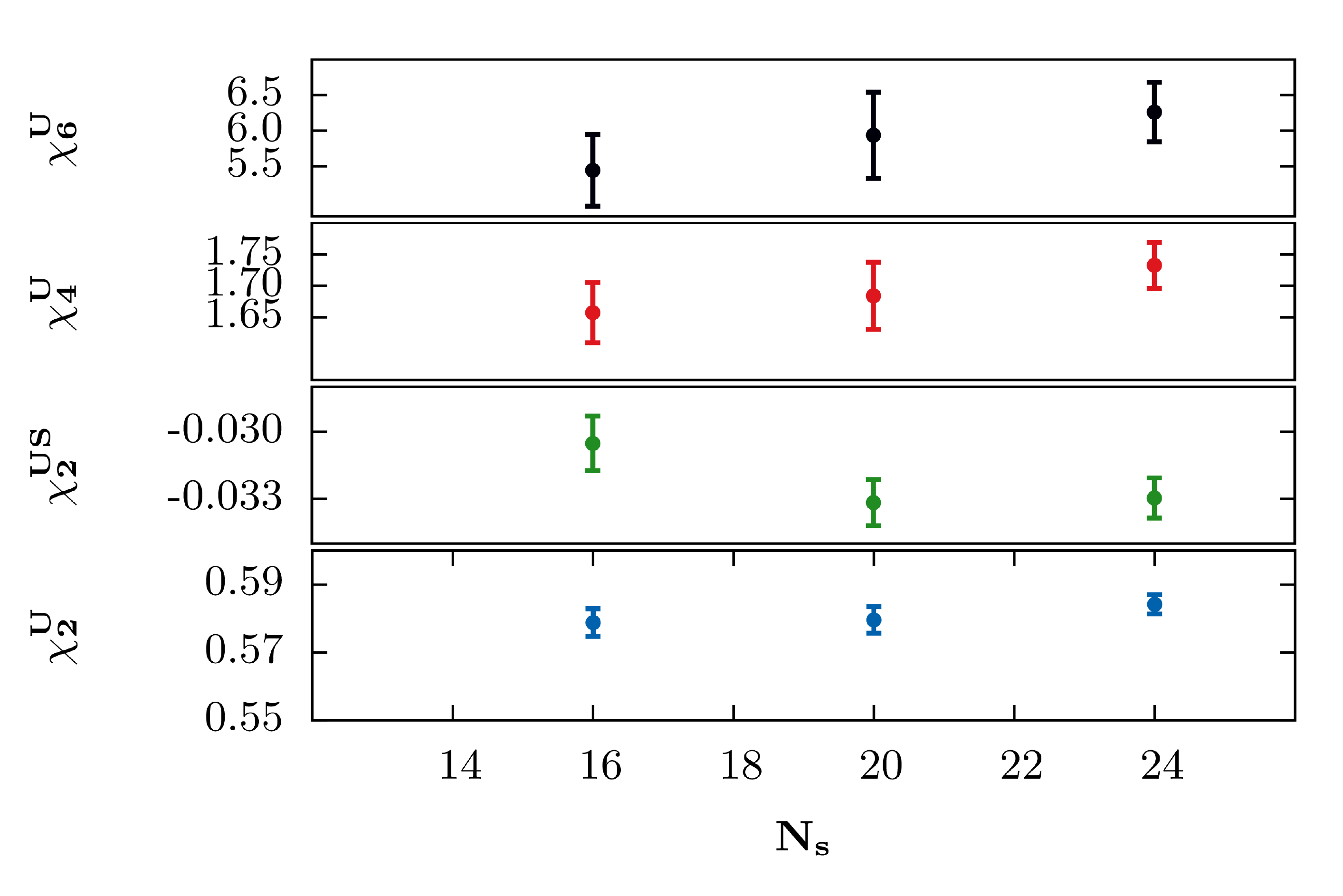}
\caption{Variation of some quark susceptibilities with the volume
  size at $T = 170$ MeV on the $N_{t} = 6$ lattice.}
\label{finite_vol170}
\end{flushright}
\end{figure}

\begin{figure}
\centering
\includegraphics[width=0.95\columnwidth]{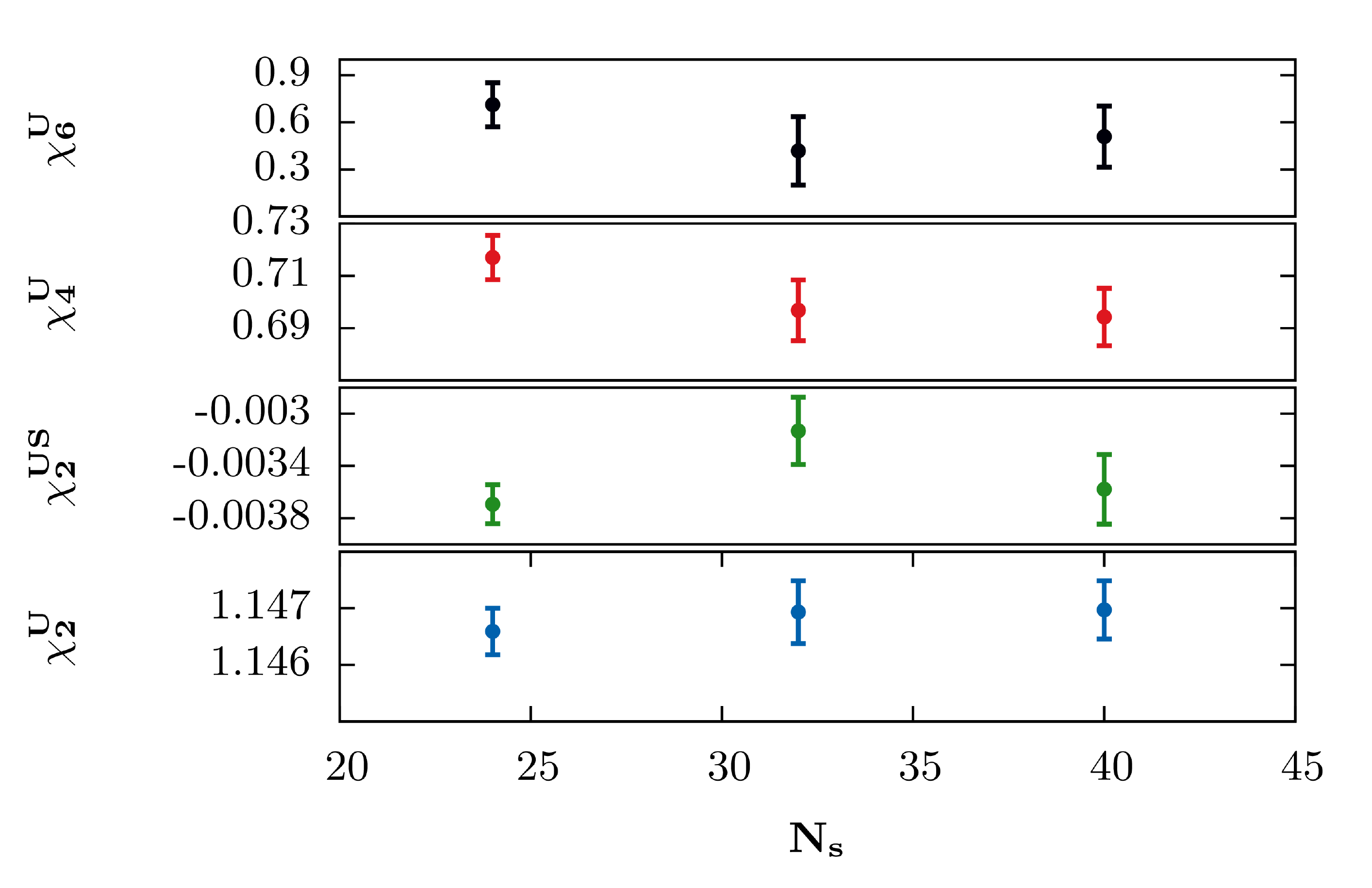}
\caption{Same as in Fig.~\ref{finite_vol170} for $T = 350$ MeV on the $N_{t} = 8$ lattice. }
\label{finite_vol_350}
\end{figure}

To investigate finite size effects, we carried out simulations on
$N_{t} = 6$ lattices for $T = 170$ MeV and on $N_{t} = 8$ lattices at
$T = 350$ MeV, considering three different values for the spatial
volume, $N_{s} = 16,20,24$ for the $N_{t} = 6$ lattice and $N_{s} =
24,32,40$ for the $N_{t} = 8$ one.  In Figs.~\ref{finite_vol170} and
\ref{finite_vol_350} our results for the up-quark and up-strange
susceptibilities are shown. The analysis indicates that no finite
volume effects are visible, within our present statistical accuracy,
when passing from aspect ratio $4$ to aspect ratio $3.3$ for $T = 170$
MeV, and from aspect ratio $5$ to aspect ratio $4$ at $350$ MeV. It is
interesting to notice a reduction of the statistical error on the
larger lattices: since the same statistics have been adopted for the
different spatial sizes, this can be related to the fact that some of
the fitted observables (quark number densities) are self-averaging,
i.e. their statistical fluctuations decreases as $1/\sqrt{V}$, while
the other are characterized by statistical fluctuations which are
independent of $V$ (the second order susceptibilities), so that, on
the whole, one expects some gain in accuracy when moving to larger
volumes.

This is visible even for the case of the sixth order susceptibilities,
whereas in the direct computation at $\mu = 0$ their determination
would be affected by a relative error growing like $\propto V^{2}$. \\

\subsection{Efficiency of the method and comparison with 
a direct determination at $\mu = 0$}
\label{sub-efficiency}

\begin{table}
\centering
\renewcommand{\arraystretch}{1.5}
\begin{tabular}{|c|c|c|c|c|}
\hline
 & \multicolumn{2}{c}{ $\mathsf{\mu = 0}$} & \multicolumn{2}{c}{  \textsf{From global fit}}   \\
\hline
$T$[MeV]  & 143 & 260 & 143 & 260  \\
\hline
$\chi_{2,0,0}$ & 0.410(12) & 1.0880(12)  & 0.4160(40) & 1.0883(8) \\
\hline
$\chi_{0,0,2}$ & 0.1862(24) & 1.0250(13) & 0.1865(15)  & 1.0255(10) \\
\hline
$\chi_{1,0,1}$ & -0.031(3) & -0.00774(57) & -0.031(1)  &  -0.00740(40) \\
\hline
$\chi_{1,1,0}$ & -0.075(8) & -0.0091(6) & -0.0680(20) & -0.0080(5) \\
\hline
$\chi_{4,0,0}$ & 1.1(8) & 0.65(1) & 1.250(70) & 0.635(20) \\
\hline
$\chi_{0,0,4}$ & 0.336(40) & 0.721(15) & 0.300(15)  & 0.710(30) \\
\hline
$\chi_{2,0,2}$ & 0.17(7) & 0.0452(35)  & 0.1195(33) & 0.0440(60) \\
\hline
$\chi_{2,2,0}$ & 0.2(3) & 0.043(4)  & 0.2924(82) & 0.038(5) \\
\hline
\end{tabular}
\caption{Comparison of results obtained for 2nd and 4th order
  susceptibilities from the global fit procedure, with the ones
  obtained from the standard computation and a statistics similar to
  that accumulated for $O(10)$ simulation points at imaginary $\mu$.
The total computational 
effort spent in the global fit is larger than
that spent in the standard case by 
a factor 10 for
$T = 143$ MeV, and 3 for $T = 260$ MeV.
}
\label{Tab_confronto}
\end{table}

At this stage we are in a position to discuss the efficiency of the
method, i.e. to compare the total computational effort in the direct
calculation and in the analytic continuation method. In
Table~\ref{Tab_confronto}, we compare results obtained for 2nd and
4th order susceptibilities, for two values of the 
temperature ($T=143,260$ MeV), from the standard method
and from the global fit, in order to test the efficiency of our method both in the confined and in the plasma phase.

In order to make a proper comparison, one must take the relative computational
effort into account.
In both cases, each measurement involved 256 random sources, however 
5 matrix inversions for each flavor were used in the standard
determination, 
in order to obtain all susceptibilities
up to order 4, and just 2 inversions in the analytic continuation case,
in order to obtain all the second order susceptibilities
involved in the global fit.
For the standard determination, we performed
measurements on 1000 configurations for $T = 143$ MeV and 2000
configurations for $T = 260$ MeV, each separated by 10 RHMC trajectories;
the relative cost\footnote{This estimate is specific
to our code implementation on the BlueGene/Q machine and could 
be different for other implementations or machines.} of each measurement
compared to each MD trajectory was about 40 for $T = 143$ MeV
and about 7 for $T = 260$ MeV.
 The determination from analytic continuation, considering
all simulation points, involved measurements on 20K configurations
for $T = 143$ MeV and 7K
configurations for $T = 260$ MeV, each separated by 10 RHMC trajectories;
the relative cost of each measurement
compared to each MD trajectory was about 16 for $T = 143$ MeV
and about 3 for $T = 260$ MeV.
Summing up, we can estimate a total computational 
effort spent in the global fit which is larger than
that spent in the standard case, 
by a factor
10 for $T = 143$ MeV, and 3 for $T = 260$ MeV.
In standard
importance sampling, error bars scale according to the inverse square
root of the sample size;
therefore, rescaling appropriately the error, we can compare the two determination at fixed machine time. 

A clear result, emerging from Table~\ref{Tab_confronto}, is that
the standard method is comparable, or even more efficient than analytic 
continuation in the deconfined phase,
for all susceptibilities up to order four. For $T = 143$ MeV,
i.e. below the pseudocritical temperature, the situation is 
quite different. Analytic continuation has still an efficiency
comparable to the standard method for second order susceptibilities,
however for fourth order susceptibilities the improvement is dramatic:
analytic continuation 
leads to an improvement which is of order 10, in terms of time machine,
for the diagonal light quark susceptibility, $\chi_{4,0,0}$, and grows
up to order 100 for the non-diagonal susceptibilities
(no significant improvement is observed, instead, for
$\chi_{0,0,4}$).

\begin{table}
\centering
\renewcommand{\arraystretch}{1.5}
\begin{tabular}{|c|c|c|c|c|}
\hline
$n_{copies}$ & $64$ & $128$ & $256$ & $512$  \\ 
\hline
$\chi_{2,0,0}$ & 0.401(37) & 0.400(19) & 0.410(12) & 0.4158(81)   \\
\hline
$\chi_{0,0,2}$ & 0.186(5) & 0.191(3) & 0.1862(24) & // \\
\hline
$\chi_{1,0,1}$ & -0.031(9) & -0.026(5) &  -0.031(3) & // \\
\hline
$\chi_{1,1,0}$ & -0.084(24) & -0.084(12) & -0.075(8) & // \\
\hline
$\chi_{4,0,0}$ & 9(7) & 3(2) & 1.1(8) & 1.05(35) \\ 
\hline
$\chi_{0,0,4}$ & 0.16(19) & 0.33(6) & 0.336(40) & // \\ 
\hline
$\chi_{2,0,2}$ & 0.12(45) & 0.10(15) & 0.17(7) & // \\
\hline
$\chi_{2,2,0}$ & 3(3) & 0.7(9) & 0.2(3) & // \\
\hline
\end{tabular}
\caption{A subset of 2nd and 4th order susceptibilities as a function of the number of random sources is shown. Data refer to simulations at $\mu=0$ and $T=143$ MeV.}
\label{tab_copie}
\end{table}

For sake of completeness, in Table~\ref{tab_copie}, we report the values of some 2nd and 4th order susceptibilities as a function of the number of random sources. Our determinations suggest that the error over 4th order cumulants decreases more sharply with respect to the 2nd order ones when increasing the number of random vectors. This different behaviour is expected, since 4th order cumulants are composed by terms which involve products of three and four traces and their uncertainty decreases more sharply as the number of random vectors is increased. For sure, by increasing sufficiently the number of random sources error saturation will occur due to the fact that gauge fluctuations dominate over random noise. However, it is possible that going from $256$  to $512$ or $1024$ 
random sources, this trend continues to be valid. Therefore, uncertainties over quark number susceptibilities determined from direct sampling  and from the global fit
could scale differently as the number of sources is increased, 
leading to a slight change in the 
efficiency comparison, which however should not change the main conclusion,
i.e. that analytic continuation gains a large factor, below $T_c$,
starting from fourth order susceptibilities, and especially for mixed ones.

\begin{figure}[]
\centering
\includegraphics[width=0.95\columnwidth]{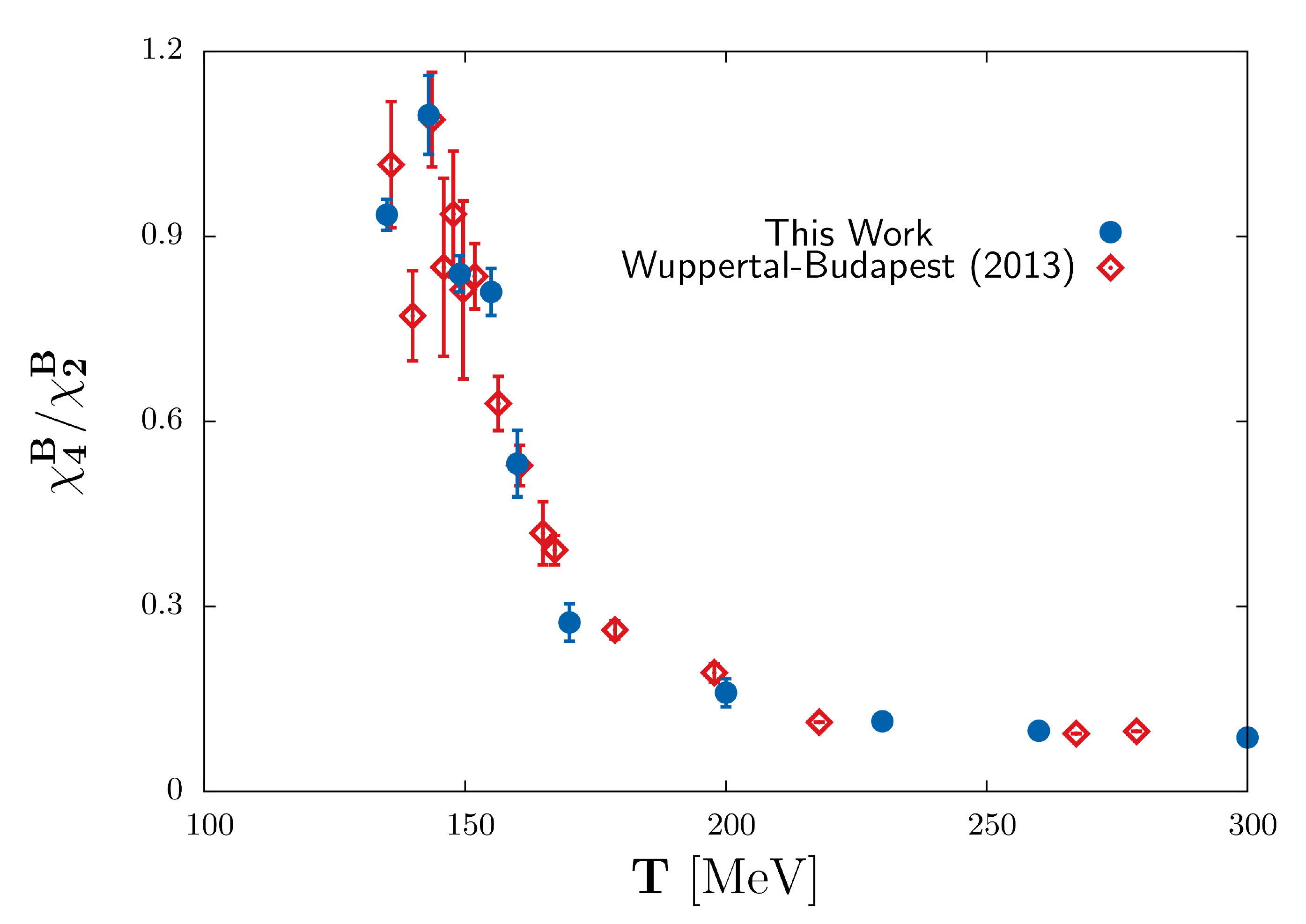}
\caption{Temperature dependence of the ratio $\chi_{4}^{B}/
  \chi_{2}^{B}$ of baryonic cumulants. Blue points correspond to our
  determinations while red points, corresponding to data obtained on a
  $N_{t} = 8$ lattice using our own discretization, are taken from
  \cite{Borsanyi:2013hza}}
\label{time_mach}
\end{figure}

Let us try to give a few possible explanations for the fact that
analytic continuation seems to be not so convenient above $T_c$.
A significant role is surely played by the fact that small eigenvalues
of the Dirac operator are strongly suppressed above $T_c$ (due to 
chiral symmetry restoration), so that,
at the same time, the multiple inversions needed in the 
standard method are less costly, and fluctuations in the noisy estimators
are suppressed; this effect is visible even below 
$T_c$, for susceptibilities involving strange quarks,
which have a larger mass, for which the gain of analytic
continuation is less marked. Another possible factor
is related to the fact that we are working with a fixed aspect
ratio, so that simulations at higher temperatures are based on
smaller physical volumes, where problems related to 
the lack of self-averaging are expected to be less severe.
Finally, in the high temperature phase analytic continuation
is surely disfavored by the reduced range of explorable chemical
potentials, due to RW or RW-like transitions: that affects
both the statistical accuracy of the global fit and, even more
important, the systematic uncertainty related to truncation
effects.
\\

\begin{figure}
\centering
\includegraphics[width=0.95\columnwidth]{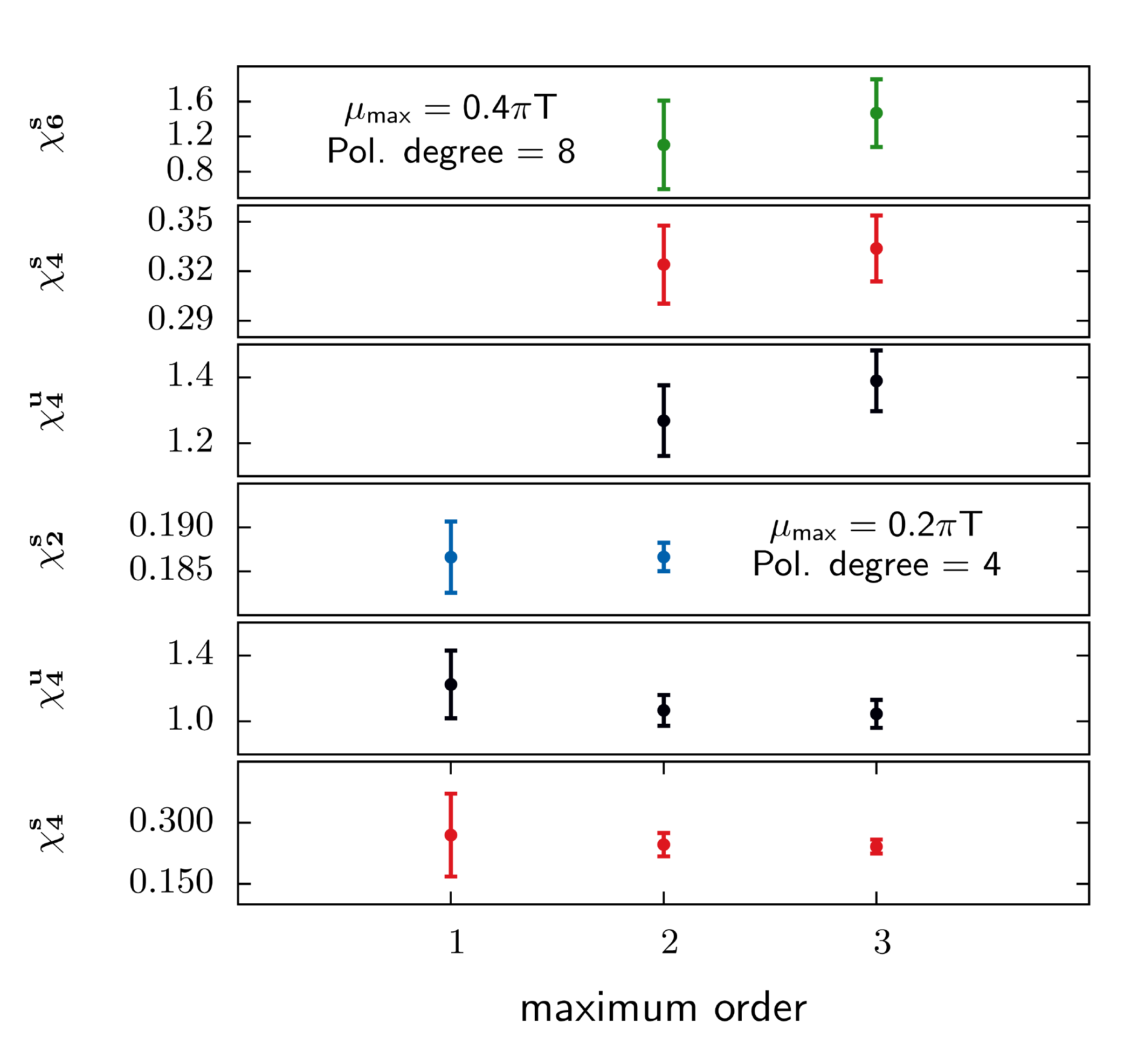}
\caption{We show how the precision attained for some susceptibilities
changes when adding more and more
   cumulants to the global fit. The first
  three graphs correspond to a global fit performed using $\mu_{max}/T
  = 0.4\pi$ and a polynomial of degree 8, whereas for the last three
  $\mu_{max}/T=0.2\pi$ and a polynomial of degree 4 has been
  used. Data refer to simulations at $T = 143$ MeV.}
\label{quality_vs_ord}
\end{figure}

Another question that we would like to answer, which regards the
optimal strategy to be followed, is whether there is any significant
gain in trying measuring also susceptibilities of order larger than 2.
In other case, like in the use of analytic continuation for the study
of $\theta$
dependence~\cite{Bonati:2015sqt,Bonati:2016tvi,Panagopoulos:2011rb},
the issue is not very important, since one can compute cumulants of
the topological charge at any chosen order with no significant
computational overhead; in this case instead, going one order further
in the measure of cumulants means adding new inversions of the Dirac
operator, with a considerable overhead.  To this purpose, we performed
trial simulations at $T = 143$ MeV, measuring all quark number
susceptibilities up to order three, and observing how errors change as
a function of the order of the susceptibilities included in the global
fit.  Some results are reported in Fig.~\ref{quality_vs_ord}.  A
remarkable improvement is achieved when adding second order
susceptibilities to the information coming from just quark number
densities: the improvement reaches up to a factor 3, in terms of error
reduction.  On the other hand, including also the third order has a
low impact, since in general only little gain is achieved.  

Finally, we would like to discuss whether the choice of equally distributed
simulation points, along the imaginary chemical potential axes, is optimal
or not. In principle, one would expect that having more simulations
where cumulants get larger contributions from higher order terms
of the expansion, i.e. at larger values of $\mu_I$, would be better,
in order to obtain more information on higher order susceptibilities. 
However, one must
consider that, in order to properly perform the analysis on the 
systematic error related to the series truncation, which has been illustrated
in the previous subsection, one needs enough determinations at
small $\mu_I$ as well. In fact, we have tried to perform the analysis
on various subsets of our simulation points, keeping more data
either in the high or in the low $\mu_I$ region and comparing the 
final error in the various cases, after normalizing it
to the total computational effort needed. The result is that
there is indeed a benefit in having more points in the 
high $\mu_I$ region when one considers just the statistical error,
however that disappears when the total error (statistical + systematic)
is taken into account, so that the choice of equally distributed points
still seems a reasonable one.\\

\begin{figure}
\centering
\includegraphics[width=0.95\columnwidth]{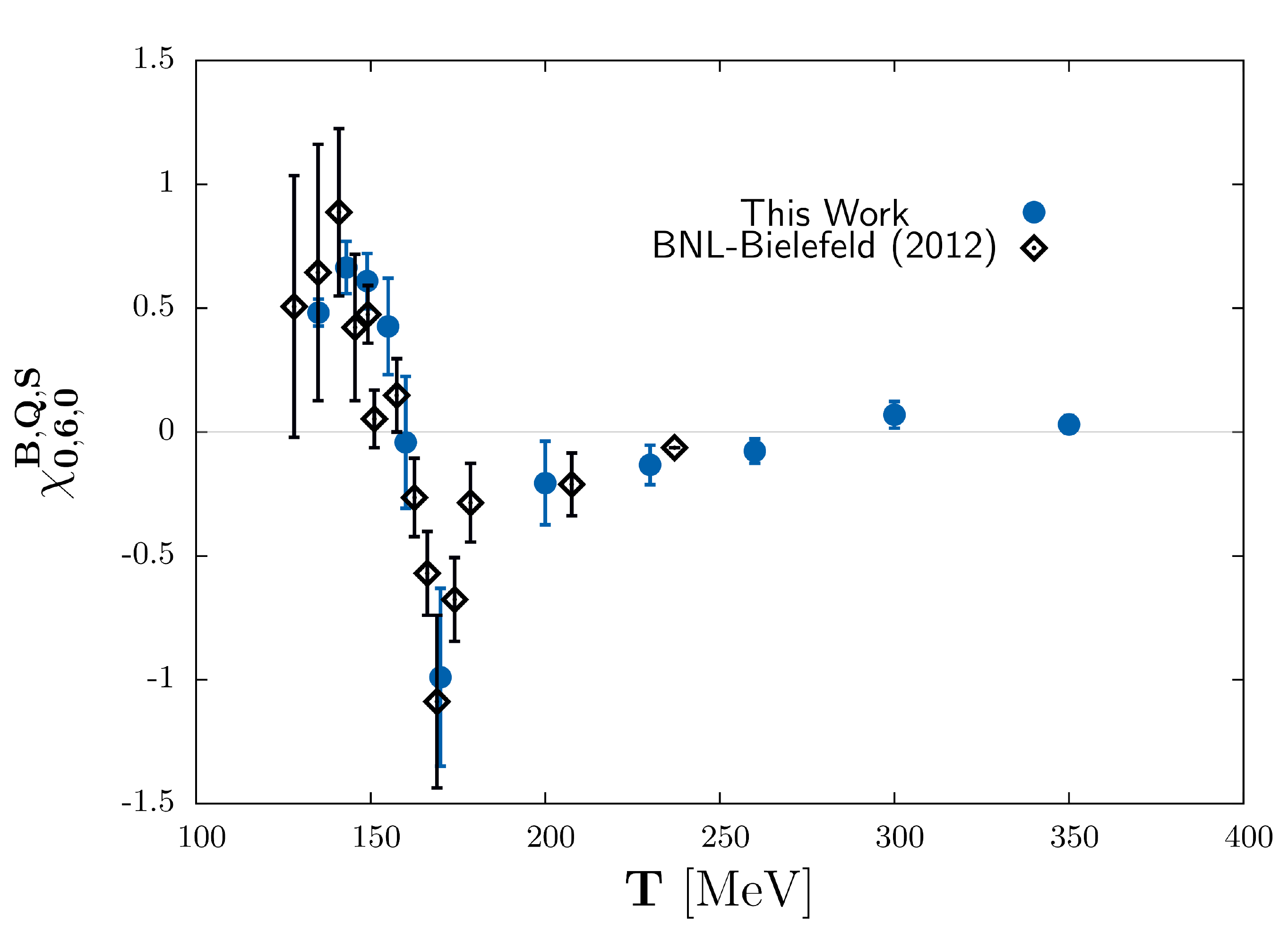}
\caption{Comparison of results on sixth order cumulant of the electric
  charge fluctuations between this work and
  Ref.~\cite{Schmidt:2012ka}. Diamonds refer to the determinations of
  Ref.~\cite{Schmidt:2012ka} obtained on a $N_{t} = 8$ lattice with
  the highly improved staggered quark (HISQ) action and almost
  physical quark masses.}
\label{Biel_0_6_0}
\end{figure}

\begin{figure}
\centering
\includegraphics[width=0.95\columnwidth]{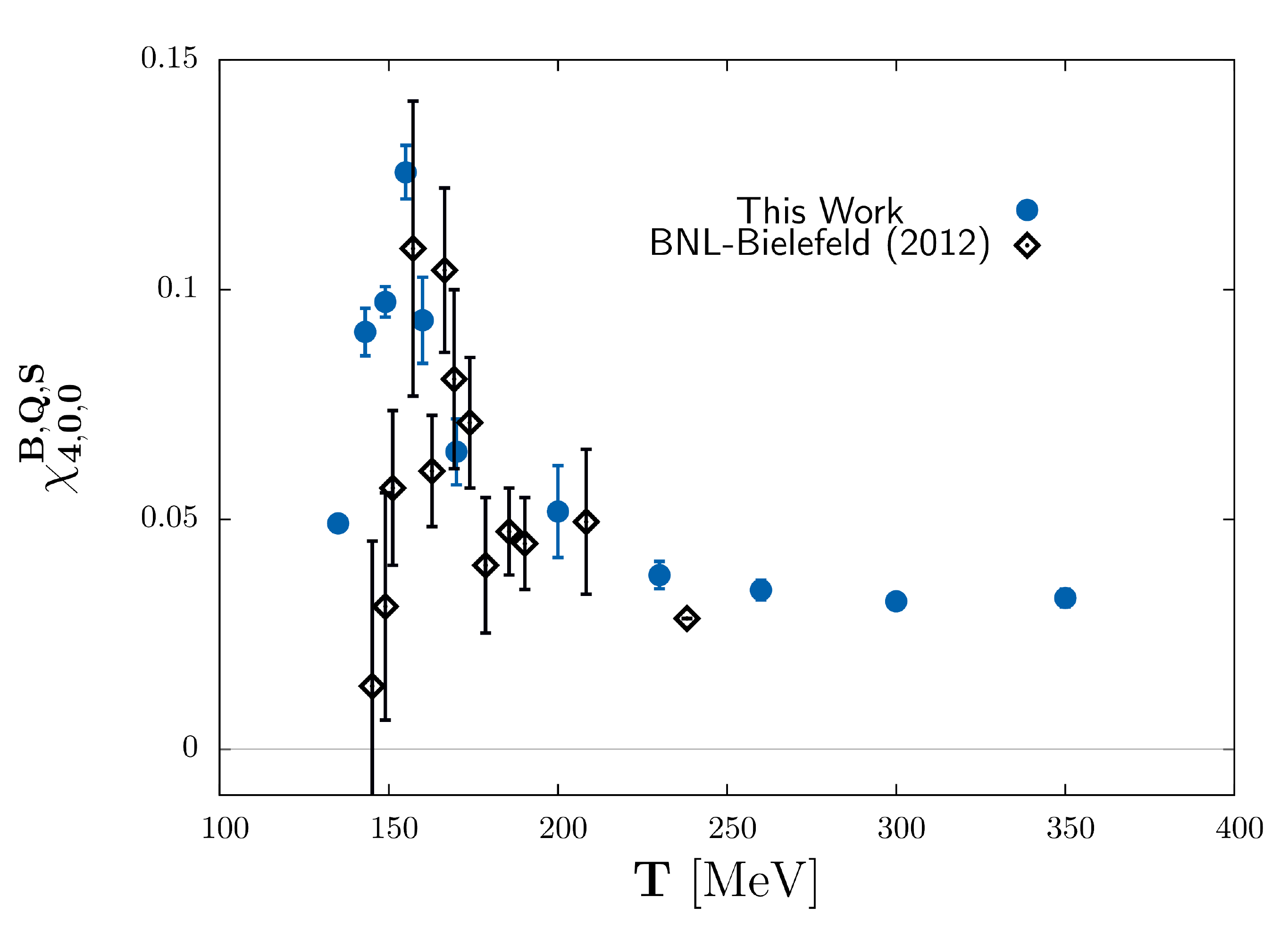}
\caption{Same as in Fig.~\ref{Biel_0_6_0} for the fourth order
  cumulant of the net baryon number fluctuation.}
\label{Biel_4_0_0}
\end{figure}

\begin{figure}
\centering
\includegraphics[width=0.95\columnwidth]{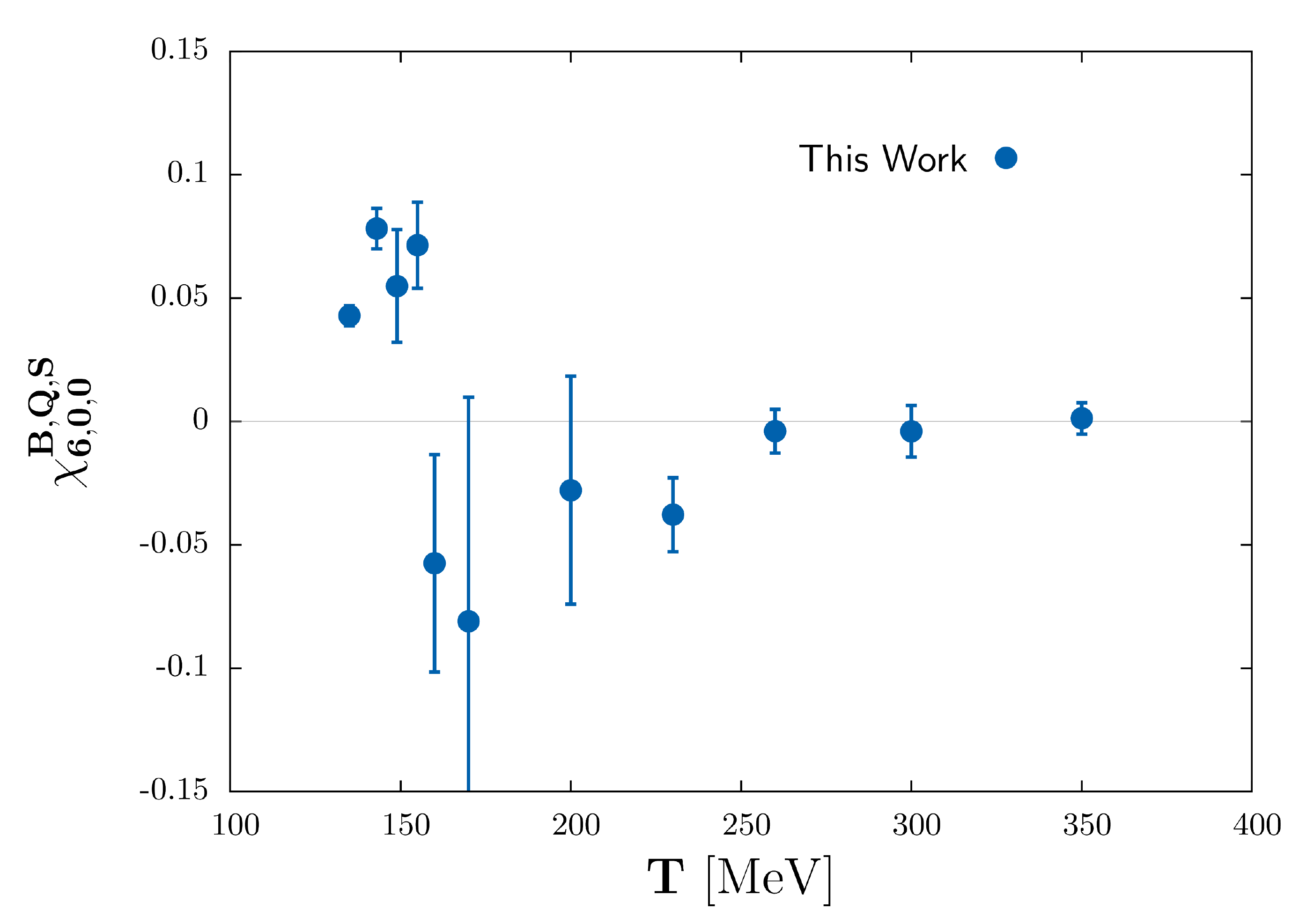}
\caption{Our determination of the sixth order cumulant of the net
  baryon number fluctuation.}
\label{Biel_6_0_0}
\end{figure}

\begin{figure}
\centering
\includegraphics[width=0.95\columnwidth]{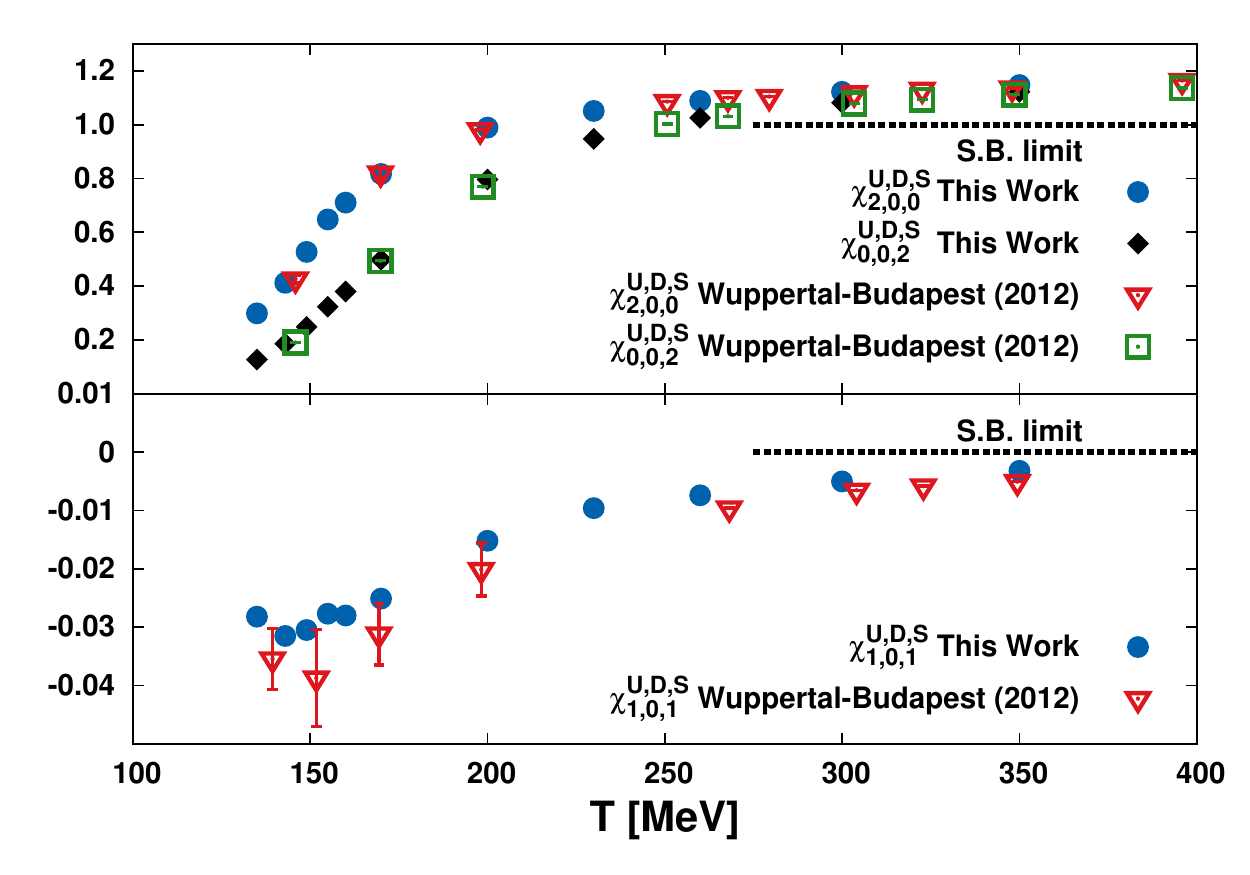}
\caption{Comparison between results on second order susceptibilities obtained in this work and in \cite{Borsanyi:2011sw} on the $N_{t} = 8$ lattice and with our own discretization.}
\label{Wup2}
\end{figure}

The complete list of susceptibilities determined on the $32^{3}\times
8$ lattice are reported in Tables~\ref{table_susc2} -
\ref{table_susc8}, while in Figs. \ref{time_mach}, \ref{Biel_0_6_0}, \ref{Biel_4_0_0},
\ref{Biel_6_0_0} and \ref{Wup2} some of those susceptibilities are
shown, as a function of $T$, and compared with results obtained by
other groups using the direct computation approach. We remind that conserved charge susceptibilities are linked to the quark number ones via linear relations (See Eq.~(\ref{defchem})). Correlations among quark number susceptibilities, as determined from the global fit procedure, 
turned out to be, in most cases, smaller than $10-15$\%. Therefore the errorbars shown in Figs. \ref{time_mach}, \ref{Biel_0_6_0}, \ref{Biel_4_0_0},
\ref{Biel_6_0_0} were computed by using Gaussian error propagation formulae.
A very good agreement is found for almost all quantities and a higher precision is
reached in our case, at least in the confined phase $T < T_c$.  Only a
small discrepancy is observed for the $\chi_{2}^{us}$ in the high
temperature regime (see Fig.~\ref{Wup2}). The source of this mismatch
can be attributed to the different aspect ratios used in the two
cases. Indeed, Ref.~\cite{Borsanyi:2011sw} adopted $N_{s}/N_{t}=3$,
while in our case we have $N_{s}/N_{t}=4$; looking at
Fig.~\ref{finite_vol_350} it is clear that finite volume effects are
still non negligible for aspect ratio 3 and for this values of the
temperature, and point exactly in the direction of the observed
discrepancy.


\subsection{An application to the search for a critical endpoint}

The obtained susceptibilities could be used for several
phenomenological analyses, like a determination of the freeze-out
line~\cite{Bazavov:2012vg,Borsanyi:2013hza,Borsanyi:2014ewa}. However,
since our results still lack of a reliable continuum extrapolation and
have been obtained essentially for one single value of $N_t$, we
prefer to postpone this to a future investigation.

\begin{figure}
\centering
\includegraphics[width=0.98\columnwidth]{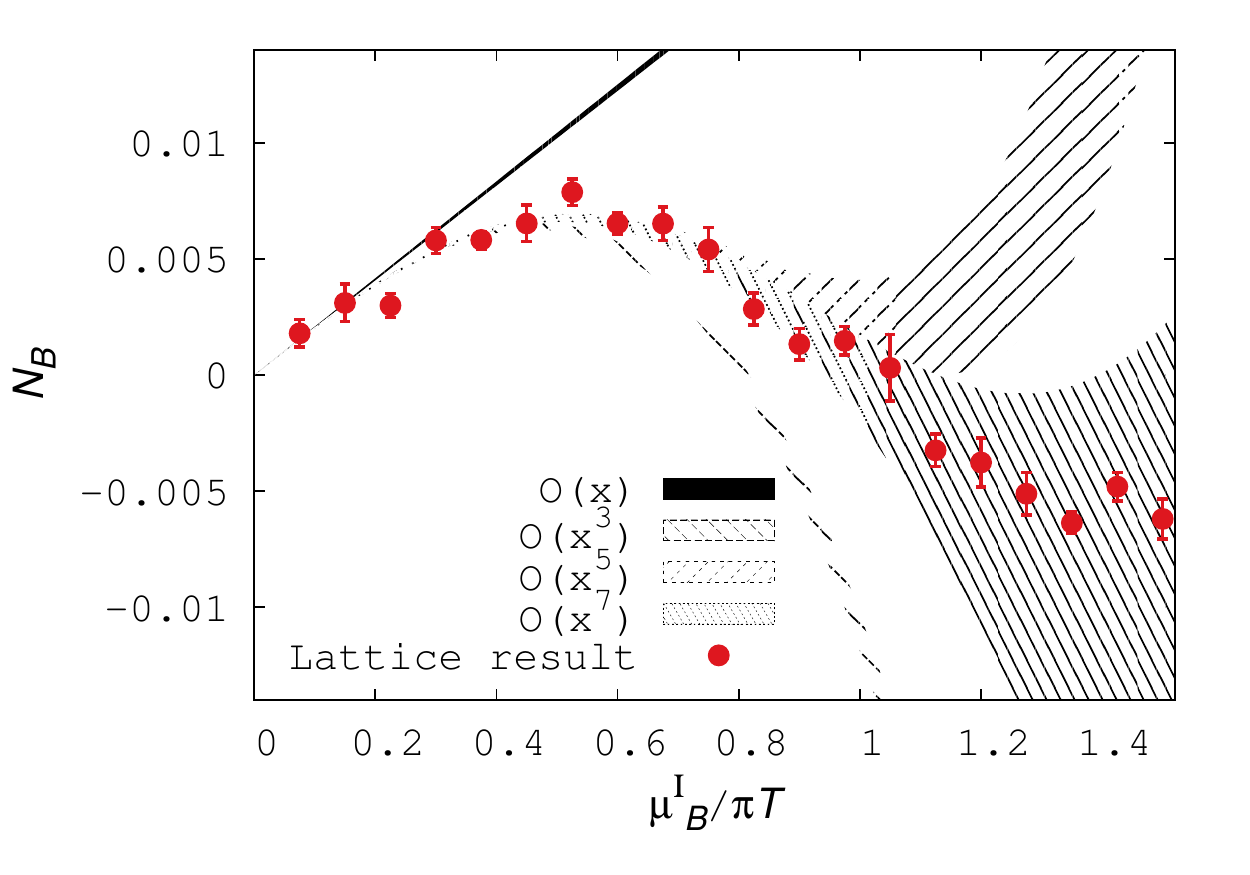}
\caption{Our lattice determination of the baryon number density as a function of the imaginary (baryon-)chemical potential is shown. Bands correspond to polynomial truncations
at various orders of the series expansion around $\mu_B = 0$. Data refer to $T=135$ MeV.  }
\label{nb_vs_mub}
\end{figure}

There is however one kind of analysis which is worth doing even
for a single value of $N_t$, and regards the
possible emergence of a critical behavior for some value of the (real)
baryon chemical potential, i.e. the existence and location of the
critical endpoint.  Indeed, high order cumulants of the net baryon
number distribution can be used to find signals of critical behavior
in the $T-\mu_{B}$ plane, following the strategy of
Refs. \cite{Gavai:2004sd, Gavai:2008zr, Datta:2012pj}. Setting
$\mu_{u}=\mu_{d}=\mu_{s}=\mu_{B}/3$ in the free energy expansion
(\ref{free_energy_expansion}) we are left with a power series in the
baryon chemical potential (see Eq.~(\ref{defchem})):
\begin{align}
\label{bar_series}
\mathcal{F}(T,\mu_{B}) = \mathcal{F}(T,0) + VT^{4}\sum_{n}\frac{\chi_{2n}^{B}}{(2n)!}\, \left(\mu_{B}/T\right)^{2n} \, .
\end{align} 
An example of the expansion is reported 
in Fig.~\ref{nb_vs_mub}, where we show, for a single value of the 
temperature, our lattice determination of the baryon number density (as a function of $\mu_{B}^{I}$), along with the various polynomial truncations of different orders
coming from its Taylor expansion around $\mu_{B}^{I}=0$.

At a second order $\mu_{B}$-driven phase transition, the free energy
develops a non-analiticity while the baryon number susceptibility
$\chi_{2}^{B}$ shows a divergence. Therefore, signals of critical
behavior can be inferred by looking for the radius of convergence of
their Taylor series. To be physical, the singularity must be placed on
the real $\mu_B$ axis, thus for the method to be effective the series
must have only positive non-null terms. In this case, estimates for
the radius of convergence of the free energy ($\rho^f$) or of the
baryon susceptibility ($\rho^\chi$) are provided by:
\begin{equation}
\label{estimators}
\rho_{n,m}^{f}= \left(
\frac{\chi_{n}^{B}/n!}{\chi_{m}^{B}/m!}\right)^{\frac{1}{(m-n)}} \,
\rho_{n,m}^{\chi}= \left(
\frac{\chi_{n}^{B}/(n-2)!}{\chi_{m}^{B}/(m-2)!}\right)^{\frac{1}{(m-n)}}
\end{equation} 
and they all coincide when the infinite $m$ and/or $n$ limit is taken.
In our case, by using the few number of coefficients at our disposal a consistent
determination of the critical endpoint requires that all the
estimators in Eq.~(\ref{estimators}) agree with each other or at least
show some signal of convergence. 
Of course, the number of terms needed to have such convergence 
is not known apriori and depends on the nature of the critical point, if it
exists. However, 
we tested the possibility of finding the critical point using this method 
by bulding up a simple statistical toy model, 
the interested reader will find more details in Appendix \ref{sec:appA}.

\begin{figure*}
\centering
\scalebox{0.40}{\includegraphics{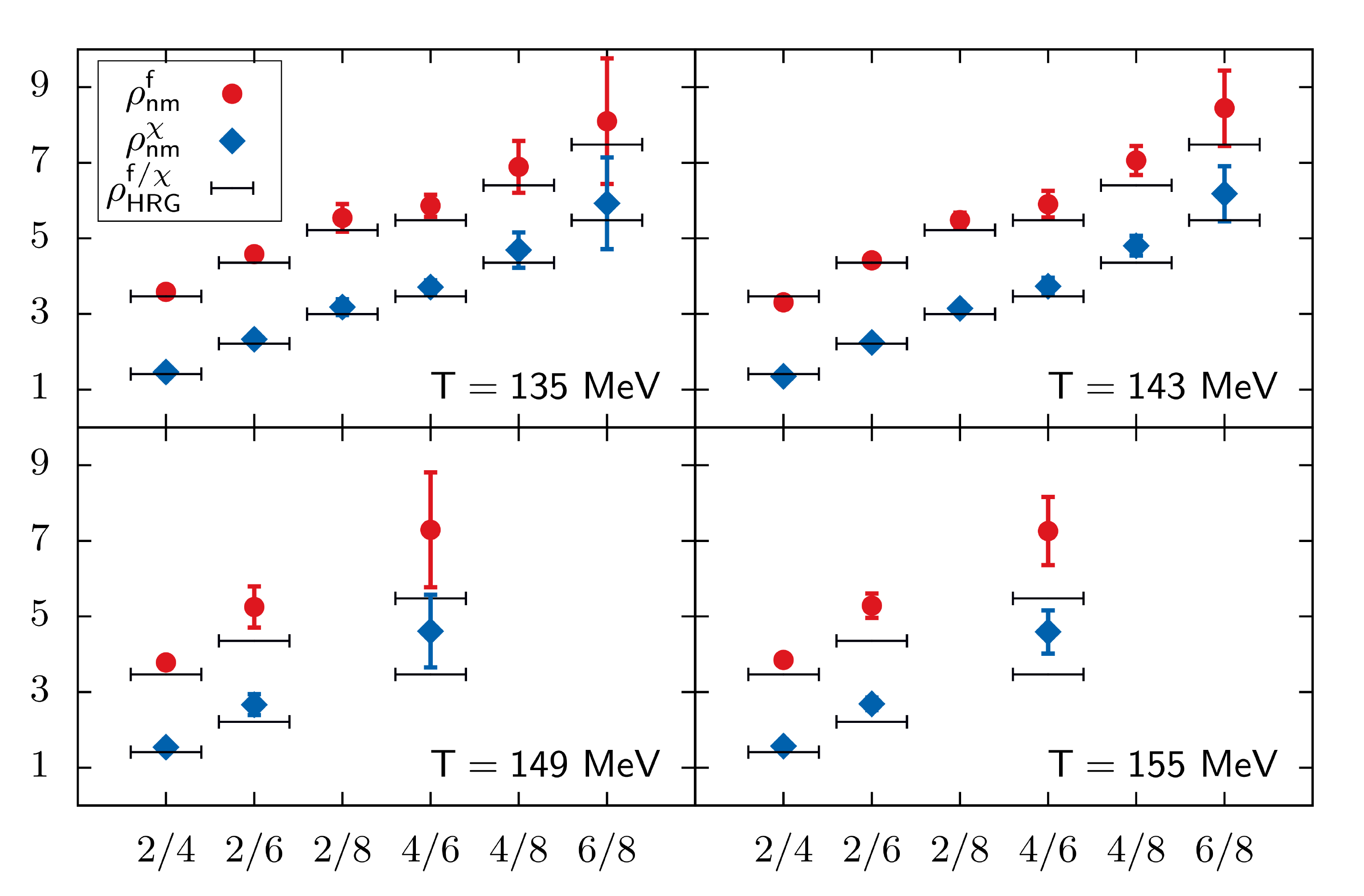}}
\caption{Radius of convergence estimates for various temperatures
  below $T_{c}$. Circles/diamonds correspond to our estimate for
  $\rho_{n,m}^{f/\chi}$ while black lines are values predicted from
  the HRG model.}
\label{panel_susc}
\end{figure*}

Since the pseudocritical line bends down for real baryon chemical
potentials, the critical endpoint, if any, is expected for
temperatures $T\le T_{c}\sim 155$ MeV. Hence, we evaluated the
estimators in Eq.~(\ref{estimators}) using the susceptibilities up to
$\chi_{8}^{B}$ for $T=135,143$ MeV, and up to $\chi_{6}^{B}$ for
$T=149,155$ MeV. For this values of temperature all the determined $\chi_{2n}^{B}$ appear to be greater than zero hence allowing for such a kind of analysis. The panels in Fig.~\ref{panel_susc} display our
determinations, where they are also compared to
the same quantities as extracted from a simple HRG model, where
\begin{equation}
F(T,\mu_{B})_{HRG} = A(T) + B(T)\, \cosh 
\left( \frac{\mu_{B}}{T} \right) 
\end{equation}
and of course the asymptotic radius of convergence is infinite. 
As it can be noticed, the estimated radii do not seem to converge to
constant values as the order increases, but rather they are in good
agreement with HRG estimates for $T < 0.95\, T_{c}$.  For $T \gtrsim
0.95\ T_c$, deviations are visible, however they correspond to
estimated radii which are larger than the HRG expectation.

\begin{figure}
\centering
\includegraphics[width=0.95\columnwidth]{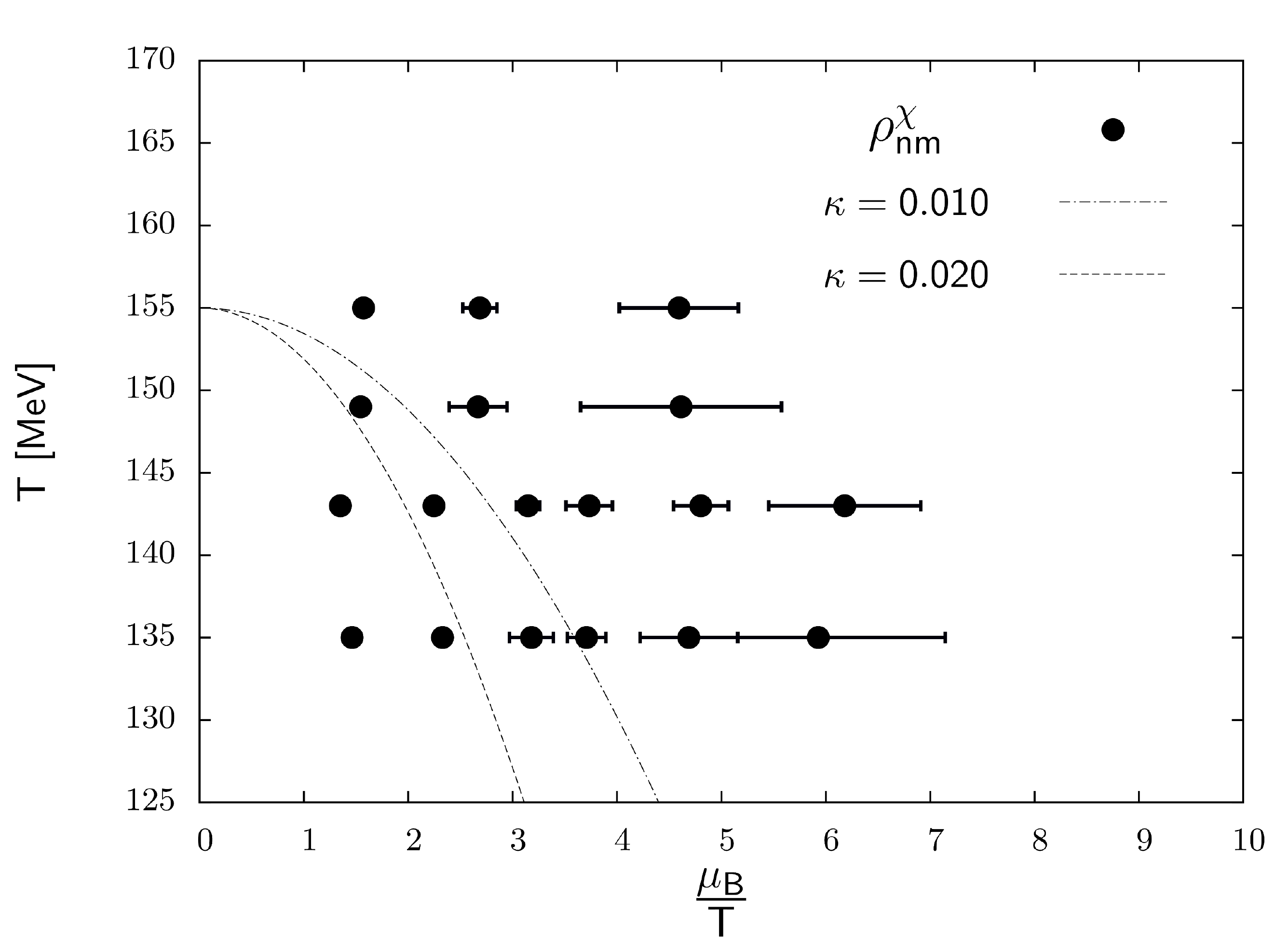}
\caption{The values of $\rho^{\chi}_{n,m}$ are shown along with the
  $\mathcal{O}(\mu^{2}_{B})$ determination for the pseudocritical
  chiral line. }
\label{pseudo}
\end{figure}

The critical endpoint should be located somewhere along the 
pseudocritical starting from $\mu_B = 0$. Therefore, it is interesting 
to report our estimated radii in the phase diagram together
with the pseudocritical line as estimated from its curvature 
at $\mu_B = 0$, i.e.
\begin{equation}
\frac{T_{c}(\mu_{B})}{T_{c}} = 1 - \kappa\left(\frac{\mu_{B}}{T_{c}}\right)^{2} + O(\mu^{4}_{B})\, . 
\end{equation}
This is shown in Fig.~\ref{pseudo}, where a range of values of
$\kappa$ is reported, going from $0.010$ to $0.020$, which roughly
corresponds to the indications from most recent lattice
determinations~\cite{Bonati:2014rfa, Bonati:2015bha, Cea:2014xva,
  Cea:2015cya, Hegde:2015tbn,Endrodi:2011gv}.  The estimated radii
rapidly exceed, as the order in the expansion grows, the position of
the estimated crossover line.

Therefore, we conclude that the present indication is that either no
criticality appears at these temperatures, or that higher order
coefficients would be required in order to be sensible to the singular
part of the free energy. We stress that to put this result on more
solid grounds, either in favor or in disfavor of a CEP at these
temperatures, several successive coefficients $\chi_{n}^{B}$ could be
required\footnote{As example, in Ref.~\cite{York:2011km} the authors tried to determine the Critical Point of the 3D Ising Model $(T_{c},H_{c}=0)$ putting an external magnetic field $H_{o}$ and then evaluating several cumulants of the free energy Taylor expansion in $\frac{(H-H_{o})}{T}$ at fixed temperature. They found that an accurate determination of the Critical Point $(H_{c} \approx 0)$ by means of radius of convergence estimates, requires the evaluation of 
at least 8 coefficients in the cumulant expansion.}.
Moreover, the same analysis should be repeated for different values of
$N_t$ to control UV cutoff effects.

\section{Discussion and Conclusions}\label{sec:conclusions}

In this work we studied $N_{f}=2+1$ QCD by means of analytic
continuation from three different imaginary chemical potentials
coupled to the up, down and strange quarks. We performed simulations
for 11 values of the temperature, using a $32^{3} \times 8$ lattice
with a stout staggered fermion discretization, the tree level Symanzik
improvement for the pure gauge part and physical quark masses.  First
and second order free energy derivatives were measured as a function
of the purely imaginary chemical potentials, and then interpolated by
means of polynomial functions in order to reconstruct the Taylor
expansion of the free energy around $\{\mu_{i}\}_{i=u,d,s} = 0$.  The
chosen trajectories in the imaginary $\{\mu_{i}\}_{i=u,d,s}$ space
(see Eq.~(\ref{chosenlines}) ensure the possibility to estimate all
kind of fluctuations and cross-correlations among conserved charges up
to order eight.  Different ranges of chemical potentials and different
polynomials have been used, in order to monitor systematic effects
related to analytic continuation.  Different spatial sizes have been
also investigated, both below and above $T_c$, obtaining as a result
that finite size effects are well under control if an aspect ratio at
least 4 is used. No systematic analysis has been performed regarding
UV cutoff effects: our results are mostly limited to $N_t = 8$ lattices
and a continuum extrapolation is postponed to a future investigation.

One of the main purposes of this study was that of checking the
efficiency of the method, as compared with the standard determination
of non-linear susceptibilities from simulations at zero chemical
potentials, and give indications about the optimal strategy to be
followed. We provided susceptibilities up to order six for $9$ values
of temperature and up to order $8$ for $T=135,143$ MeV, where the
extended range of measurements at imaginary $\mu$ $(\mu_{I,max}=0.8\,
\pi/T)$ allowed us to fit polynomials up to order ten. Our results are
in good agreement with previous standard determinations.
Regarding efficiency, we obtained that analytic continuation can 
lead to a significant improvement below the pseudocritical 
temperature $T_c$: in term of computational cost, this 
improvement is of order 10 for fourth order diagonal light 
quark susceptibilities, and goes up to a factor 100
for non-diagonal ones; we could not make a direct test
for higher order susceptibilities, for which 
the improvement is expected to be even larger.

On the contrary, analytic continuation does not reveal to be 
a competitive strategy above $T_c$. One possible reason
is related to the restoration of chiral symmetry, which 
causes a significant reduction in the statistical fluctuations present
in the noisy estimators and in the numerical cost
of matrix inversions: both these factors go in the direction
of a strong improvement in the standard determination. Another
possible reason is related to 
the reduced range of explorable imaginary chemical
potentials, due to RW or RW-like transitions, which affects
both the statistical accuracy of the global fit and
the systematic uncertainty related to truncation
effects.

The precision reached below $T_c$ allowed us to perform an analysis
regarding the possible location of the CEP. We evaluated cumulants of
the net baryon number fluctuations for four values of $T\leq T_{c}$:
up to $\chi_{8}^{B}$ for $T=135,143$ MeV and up to $\chi_{6}^{B}$ for
$T=149,155$ MeV. Various estimators of the radius of convergence of
the Taylor expansion, both for the free energy and for the baryon
susceptibility, have been considered.  We did not observe any signal
of convergence of the estimated radii, and for $T \lesssim 0.95\,
T_{c}$ the estimates are consistent with a HRG-like
behavior. Moreover, the estimated radii go well beyond the estimated
location of the pseudocritical line as the order of the estimator
increases.  We retain that this result could be
interpreted in two possible ways:\\ {\em i)} No critical endpoint
exists, at least for the discretization of QCD adopted in the present
study, and within the explored range of temperatures;\\ {\em ii)}
$6$th/$8$th order baryon number susceptibilities are still not
sufficient to be sensitive to the singular part of the free energy;
moreover the critical endpoint could be located for large chemical
potentials, for which present lattice methods, which work well for
small values of $\mu_{B}/T$, are not well suited.

Finally, let us discuss how our results and the method could be
improved in future studies, especially in view of an extension to
finer lattices, in order to perform a continuum extrapolation.  An
outcome of our investigation is that the information on quadratic
susceptibilities allows to achieve a much better overall accuracy on
the global fit, and a significant improvement with respect to the
measurement of quark number densities only, whereas the inclusion of
third order cumulants does not have a significant impact. This is
important in order to define a strategy aimed at computing higher
order susceptibilities. Indeed, looking at
Figs.~\ref{rank}-\ref{rank3}, we see that the number and the order of
non-linear susceptibilities which one is able to determine can be
increased by either increasing the number of measured
susceptibilities, or by increasing the number of trajectories in the
chemical potentials space along which the simulation points are taken.
However, in view of the difficulty in adding statistically significant
information by measuring third order susceptibilities, the suggested
strategy for the future is to measure directly free energy derivatives
up to order two and to add more trajectories of simulated points.  For
instance, adding two more lines to Eq.~(\ref{chosenlines}) (see
Fig.~\ref{rank3}), corresponding to an increase in computational
effort of about 1/3, would allow to completely determine
susceptibilities up to order 12.

\begin{center}
\begin{table*}
\renewcommand{\arraystretch}{2.5}
\begin{adjustbox}{max width=\textwidth}
\begin{tabular}{ c|c|c|c|c|c|c|c|c|c|c|c}
\hline
$T$ [MeV] & 135 & 143 & 149 & 155 & 160 & 170 & 200 & 230 & 260 & 300 & 350 \\ 
\hline
$\chi_{0,0,2}$ & 	0.12770(90) & 	0.1865(15) & 	0.2485(10) & 	0.3230(20) & 	0.3800(20) & 	0.4980(30) & 	0.7960(20) & 	0.9485(15) & 	1.0255(10) & 	1.0815(10) & 	1.12250(70) \\ 
\hline 
$\chi_{1,0,1}$ & 	-0.02820(50) & 	-0.031(1) & 	-0.03050(70) & 	-0.02780(80) & 	-0.0285(10) & 	-0.0251(11) & 	-0.01520(70) & 	-0.00960(40) & 	-0.00740(40) & 	-0.00500(40) & 	-0.00320(30) \\ 
\hline 
$\chi_{1,1,0}$ & 	-0.0698(15) & 	-0.0680(20) & 	-0.0687(18) & 	-0.0566(16) & 	-0.0540(25) & 	-0.0462(20) & 	-0.0187(10) & 	-0.01130(60) & 	-0.00800(50) & 	-0.00510(30) & 	-0.00343(35) \\ 
\hline 
$\chi_{2,0,0}$ & 	0.3020(30) & 	0.4160(40) & 	0.5275(25) & 	0.6480(40) & 	0.7080(40) & 	0.8170(60) & 	0.9888(15) & 	1.0515(12) & 	1.08830(80) & 	1.12200(60) & 	1.14700(60) \\ 
\hline 

\end{tabular}
\end{adjustbox}
\caption{Table of second order susceptibilities obtained from
  polynomial fits. Errors are calculated taking into account both
  statistical uncertainties and systematic effects.}
\label{table_susc2}
\end{table*}
\end{center}
\begin{center}
\begin{table*}
\renewcommand{\arraystretch}{2.5}
\begin{adjustbox}{max width=\textwidth}
\begin{tabular}{ c|c|c|c|c|c|c|c|c|c|c|c}
\hline
$T$ [MeV] & 135 & 143 & 149 & 155 & 160 & 170 & 200 & 230 & 260 & 300 & 350 \\ 
\hline
$\chi_{0,0,4}$ & 	0.195(10) & 	0.300(15) & 	0.411(13) & 	0.470(30) & 	0.67(10) & 	0.72(12) & 	0.810(80) & 	0.75(5) & 	0.710(30) & 	0.640(35) & 	0.690(30) \\ 
\hline 
$\chi_{1,0,3}$ & 	-0.0266(32) & 	-0.0188(60) & 	-0.0200(70) & 	0.0060(80) & 	0.000(40) & 	0.038(24) & 	-0.023(25) & 	0.010(10) & 	0.007(10) & 	0.0020(60) & 	0.0080(70) \\ 
\hline 
$\chi_{1,1,2}$ & 	0.0080(20) & 	0.010(4) & 	0.02(1) & 	0.0210(60) & 	0.013(15) & 	0.010(15) & 	0.025(25) & 	0.0022(64) & 	0.0000(50) & 	-0.0065(45) & 	-0.0020(40) \\ 
\hline 
$\chi_{2,0,2}$ & 	0.0790(25) & 	0.1195(33) & 	0.160(10) & 	0.180(20) & 	0.177(20) & 	0.150(20) & 	0.084(20) & 	0.025(10) & 	0.0440(60) & 	0.0330(50) & 	0.0250(50) \\ 
\hline 
$\chi_{2,1,1}$ & 	0.0084(20) & 	0.0090(20) & 	0.0075(75) & 	0.021(8) & 	0.010(8) & 	0.0028(83) & 	0.000(10) & 	0.0140(40) & 	-0.0020(30) & 	0.0012(16) & 	0.0030(30) \\
\hline 
$\chi_{2,2,0}$ & 	0.2115(80) & 	0.2924(82) & 	0.364(12) & 	0.430(20) & 	0.310(35) & 	0.140(40) & 	0.081(14) & 	0.0434(66) & 	0.0380(50) & 	0.0280(45) & 	0.026(8) \\ 
\hline 
$\chi_{3,0,1}$ & 	-0.0060(40) & 	0.010(10) & 	0.0208(73) & 	0.050(15) & 	-0.01(3) & 	0.000(25) & 	0.020(20) & 	0.0080(70) & 	0.0040(60) & 	0.0037(51) & 	0.0050(50) \\ 
\hline 
$\chi_{3,1,0}$ & 	-0.0160(45) & 	0.010(9) & 	0.0230(90) & 	0.081(15) & 	0.004(45) & 	-0.035(25) & 	0.005(15) & 	0.010(10) & 	0.0064(54) & 	0.0018(28) & 	-0.0009(38) \\ 
\hline 
$\chi_{4,0,0}$ & 	0.850(20) & 	1.250(70) & 	1.410(40) & 	1.55(15) & 	1.30(20) & 	0.840(70) & 	0.620(60) & 	0.590(30) & 	0.635(20) & 	0.700(20) & 	0.710(25) \\ 
\hline 

\end{tabular}
\end{adjustbox}
\caption{Same as in 
Table~\ref{table_susc2} for the fourth order susceptibilities}
\label{table_susc4}
\end{table*}
\end{center}
\begin{center}
\begin{table*}
\renewcommand{\arraystretch}{2.5}
\begin{adjustbox}{max width=\textwidth}
\begin{tabular}{ c|c|c|c|c|c|c|c|c|c|c|c}
\hline
$T$ [MeV] & 135 & 143 & 149 & 155 & 160 & 170 & 200 & 230 & 260 & 300 & 350 \\ 
\hline
$\chi_{0,0,6}$ & 	0.60(15) & 	0.75(10) & 	1.00(50) & 	0.69(18) & 	1.98(87) & 	1.56(98) & 	-1.1(21) & 	0.43(56) & 	0.29(36) & 	-1.25(75) & 	-0.10(33) \\ 
\hline 
$\chi_{1,0,5}$ & 	-0.017(17) & 	0.080(30) & 	-0.060(90) & 	0.30(35) & 	-0.19(43) & 	0.63(50) & 	-1.00(70) & 	0.09(17) & 	0.06(11) & 	0.01(11) & 	0.043(87) \\ 
\hline 
$\chi_{1,1,4}$ & 	-0.005(15) & 	-0.003(12) & 	0.049(42) & 	0.045(60) & 	0.00(23) & 	0.06(32) & 	0.5(10) & 	-0.067(92) & 	-0.018(60) & 	-0.133(67) & 	-0.035(44) \\ 
\hline 
$\chi_{2,0,4}$ & 	0.080(20) & 	0.142(21) & 	0.25(12) & 	0.16(10) & 	0.11(24) & 	-0.14(38) & 	0.43(60) & 	-0.25(35) & 	-0.076(64) & 	-0.190(82) & 	-0.020(48) \\ 
\hline 
$\chi_{2,1,3}$ & 	0.010(10) & 	0.0112(90) & 	-0.10(10) & 	0.019(26) & 	0.09(13) & 	-0.20(36) & 	-0.50(23) & 	0.05(15) & 	-0.007(31) & 	0.052(33) & 	0.030(22) \\ 
\hline 
$\chi_{2,2,2}$ & 	0.0050(50) & 	0.0200(80) & 	0.000(70) & 	-0.05(10) & 	0.10(20) & 	-0.110(70) & 	0.02(14) & 	-0.032(34) & 	-0.010(19) & 	-0.044(25) & 	-0.015(13) \\ 
\hline 
$\chi_{3,0,3}$ & 	-0.0160(60) & 	0.023(13) & 	-0.002(15) & 	0.005(20) & 	0.082(42) & 	-0.29(23) & 	1.7(1.0) & 	0.70(70) & 	0.001(97) & 	0.012(43) & 	0.085(68) \\ 
\hline 
$\chi_{3,1,2}$ & 	0.0100(50) & 	0.0070(80) & 	0.050(50) & 	-0.005(13) & 	-0.037(53) & 	-0.062(69) & 	0.03(10) & 	-0.015(26) & 	0.019(16) & 	-0.030(19) & 	-0.011(12) \\ 
\hline 
$\chi_{3,2,1}$ & 	0.010(10) & 	-0.002(20) & 	0.000(13) & 	0.035(25) & 	-0.180(50) & 	-0.130(50) & 	0.027(58) & 	0.018(18) & 	-0.017(12) & 	0.009(10) & 	0.0006(81) \\ 
\hline 
$\chi_{3,3,0}$ & 	0.097(50) & 	0.035(40) & 	0.11(10) & 	-0.46(17) & 	-0.53(18) & 	-0.10(14) & 	0.050(70) & 	-0.001(23) & 	0.020(23) & 	0.015(16) & 	0.0006(81) \\ 
\hline 
$\chi_{4,0,2}$ & 	0.127(21) & 	0.190(40) & 	0.30(15) & 	0.230(50) & 	-0.21(27) & 	-0.46(31) & 	-0.32(51) & 	-0.23(12) & 	-0.058(52) & 	0.120(83) & 	-0.041(36) \\ 
\hline 
$\chi_{4,1,1}$ & 	0.010(10) & 	0.000(15) & 	-0.008(16) & 	-0.003(21) & 	0.040(87) & 	-0.083(93) & 	0.08(10) & 	-0.00(10) & 	-0.013(17) & 	-0.005(16) & 	0.014(13) \\ 
\hline 
$\chi_{4,2,0}$ & 	0.325(50) & 	0.460(80) & 	0.45(10) & 	0.460(40) & 	-0.36(14) & 	-1.00(30) & 	-0.14(11) & 	-0.080(29) & 	0.011(18) & 	-0.048(20) & 	-0.007(15) \\ 
\hline 
$\chi_{5,0,1}$ & 	0.075(75) & 	0.19(15) & 	0.00(20) & 	0.20(15) & 	-0.15(43) & 	0.00(50) & 	-0.06(46) & 	-0.32(11) & 	0.090(99) & 	0.047(85) & 	0.076(64) \\ 
\hline 
$\chi_{5,1,0}$ & 	0.140(20) & 	0.260(50) & 	0.250(70) & 	0.320(90) & 	0.18(33) & 	-0.80(35) & 	-0.27(27) & 	-0.077(66) & 	0.067(42) & 	0.002(41) & 	-0.078(34) \\ 
\hline 
$\chi_{6,0,0}$ & 	4.0(5) & 	6(1) & 	4.67(43) & 	3.7(2.0) & 	-1.8(1.5) & 	-4.1(1.5) & 	-2.3(1.4) & 	-0.94(42) & 	0.14(28) & 	0.96(25) & 	0.42(21) \\ 
\hline 

\end{tabular}
\end{adjustbox}
\caption{Same as in Table~\ref{table_susc2} for the sixth order susceptibilities}
\label{table_susc6}
\end{table*}
\end{center}
\begin{center}
\begin{table*}
\renewcommand{\arraystretch}{2.5}
\begin{adjustbox}{max width=\textwidth}
\begin{tabular}{ |c|c|c|}
\hline
$T$ [MeV] & 135 & 143  \\ 
\hline
$\chi_{0,0,8}$ & 	3.0(1.5) & 	2.5(1.5) \\ 
\hline 
$\chi_{1,0,7}$ & 	0.50(50) & 	0.90(35) \\ 
\hline 
$\chi_{1,1,6}$ & 	-0.15(15) & 	0.00(20) \\ 
\hline 
$\chi_{2,0,6}$ & 	-0.05(15) & 	0.25(13) \\ 
\hline 
$\chi_{2,1,5}$ & 	0.030(30) & 	0.00(10) \\ 
\hline 
$\chi_{2,2,4}$ & 	0.000(40) & 	0.031(21) \\ 
\hline 
$\chi_{3,0,5}$ & 	0.000(20) & 	-0.07(11) \\ 
\hline 
$\chi_{3,1,4}$ & 	0.000(20) & 	0.002(15) \\ 
\hline 
$\chi_{3,2,3}$ & 	-0.025(35) & 	0.025(17) \\ 
\hline 
$\chi_{3,3,2}$ & 	0.000(25) & 	-0.007(21) \\ 
\hline 
$\chi_{4,0,4}$ & 	0.10(10) & 	0.100(70) \\ 
\hline 
$\chi_{4,1,3}$ & 	0.000(50) & 	0.050(50) \\ 
\hline 
$\chi_{4,2,2}$ & 	-0.0050(50) & 	0.070(45) \\ 
\hline 
$\chi_{4,3,1}$ & 	0.00(5) & 	-0.020(40) \\ 
\hline 
$\chi_{4,4,0}$ & 	0.35(10) & 	0.60(15) \\ 
\hline 
$\chi_{5,0,3}$ & 	0.00(10) & 	0.15(20) \\ 
\hline 
$\chi_{5,1,2}$ & 	0.025(75) & 	-0.050(50) \\ 
\hline 
$\chi_{5,2,1}$ & 	-0.020(40) & 	0.050(60) \\ 
\hline 
$\chi_{5,3,0}$ & 	0.150(60) & 	0.26(14) \\ 
\hline 
$\chi_{6,0,2}$ & 	0.30(12) & 	0.35(10) \\ 
\hline 
$\chi_{6,1,1}$ & 	0.00(30) & 	0.00(10) \\ 
\hline 
$\chi_{6,2,0}$ & 	0.70(20) & 	0.95(30) \\ 
\hline 
$\chi_{7,0,1}$ & 	0.75(75) & 	0.60(80) \\ 
\hline 
$\chi_{7,1,0}$ & 	0.35(25) & 	0.70(26) \\ 
\hline 
$\chi_{8,0,0}$ & 	20(4) & 	35(10) \\ 
\hline 

\end{tabular}
\end{adjustbox}
\caption{Same as in 
Table~\ref{table_susc2} for the eighth order susceptibilities}
\label{table_susc8}
\end{table*}
\end{center}

\acknowledgments

We thank Claudio Bonati, Philippe de Forcrand and Ettore Vicari
for useful discussions.
FS received funding from the European Research Council under the European
Community Seventh Framework Programme (FP7/2007-2013) ERC grant agreement No
279757.  
Simulations have been performed on the BlueGene/Q Fermi at CINECA (Projects
Iscra-B/RENQCD and INF16\_npqcd).

\appendix
\section{A simple statistical toy model}\label{sec:appA}

We tested the possibility of finding, by the radius of convergence estimate method, the location of a 
critical point, by using a simple test function 
(which plays the role of the baryon number density) with a non-analiticity
located at real chemical potential. We sampled this function and
its first derivative on the imaginary $\mu$ axis, by adding a statistical
Gaussian noise to the function values, in order to obtain
data points with statistical errors, then trying to reconstruct
the Taylor expansion around $\mu=0$ by means of a polynomial 
interpolation to the sampled data, adopting the same procedure
for the estimate of statistical and systematic uncertainties
adopted for the real QCD data.

\begin{figure}[h]
\centering
\includegraphics[width=0.95\columnwidth]{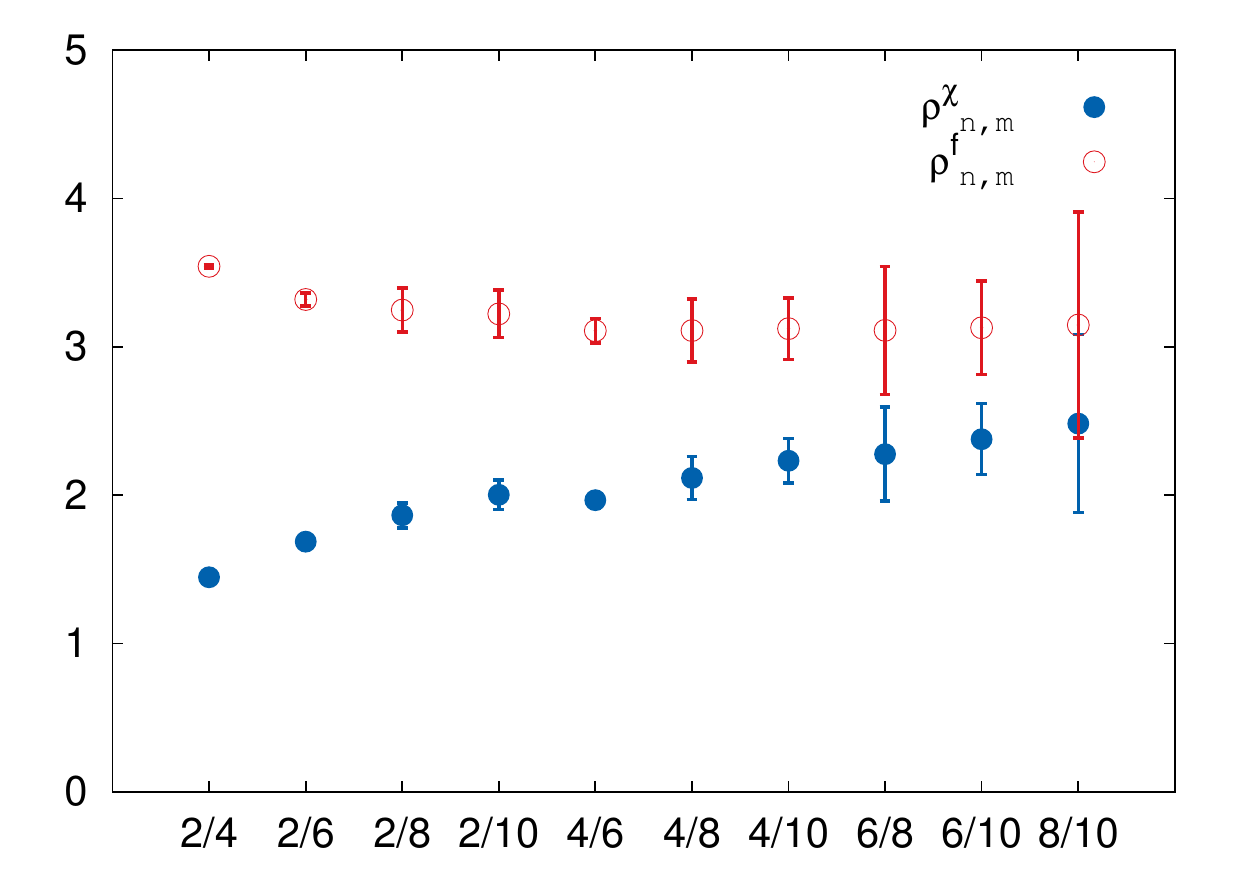}
\caption{Radius of convergence estimates for our statistical toy model. Filled points represent the radius of convergence estimates for the Taylor expansion of the susceptibility while the unfilled ones the estimates for the Taylor expansion of the free energy (See Eq.~\ref{estimators}). }
\label{test_func_est}
\end{figure}

We used as a test function 

$$
n(\mu) = \frac{\mu}{\mu_c^{2} - \mu^{2}}
$$
\\
with $\mu_c = 2.5$.
This function and its first 
derivative, which plays the role of the second order baryon susceptibility,
were sampled in the range $0\leq \mu_{I} = {\rm Im} (\mu) \leq 1.5$. 
To determine systematic errors we exactly followed the guidelines
used for quark number susceptibilities and fitted the sampled data
with polynomials up to order 12. As for the quark number susceptibilities
the very last term of the highest order polynomial we used in the best 
fit procedure was not considered, because it might have a large 
uncontrolled bias due to truncation effects. Susceptibilities up to 
order 10 were then used to compute the estimators in Eq. \ref{estimators}. 
Results are shown in Fig. \ref{test_func_est}. 

As it is clear from the figure, the estimators 
seem to convergence to the correct value of $\mu_{c}=2.5$ with the 
estimators $\rho_{n,m}^{\chi}$ showing a faster convergence with respect 
to the $\rho_{n,m}^{f}$'s. In spite of the simplicity of the statistical 
model, the important outcome is that it seems at least reasonable 
to perform, in the case of QCD, such a kind of analysis, 
even though just a few number of 
susceptibilities are known. Of course, as we have already emphasized, the 
actual number of needed terms will depend on the particular critical behavior.

\end{document}